\documentclass[final]{jfm}
\usepackage{graphicx,amssymb,amsmath,upgreek,bm}
\usepackage{dcolumn}
\usepackage{hyperref}
\usepackage{natbib}
\usepackage{rotate}
\usepackage[dvipsnames]{xcolor}
\usepackage{soul}
\usepackage{wrapfig}
\usepackage{latexsym,epsfig}
\usepackage[latin1]{inputenc}
\usepackage[T1]{fontenc}
\usepackage{dcolumn}

\newcommand{\del}{\bm{\delta}}
\newcommand{\dd}{\bm{d}}
\newcommand{\pos}{\bm{r}}
\newcommand{\vit}{\bm{u}}
\newcommand{\vitbase}{\bm{u}^{\left(0\right)}}

\newcommand{\qvit}{{\bf v}}
\newcommand{\Nab}{\bm{\nabla}}

\newcommand{\Ron}{\varepsilon}
\newcommand{\Ek}{\mbox{\textit{Ek}}}

\newcommand{\rmd}{\mathrm{d}}
\newcommand{\rmi}{\mathrm{i}}
\newcommand{\rme}{\mathrm{e}}
\newcommand{\cc}{\mathrm{c.c.}}
\renewcommand{\Imag}{\mbox{Im}}
\renewcommand{\Real}{\mbox{Re}}

\newcommand{\ex}{\bm{\hat{x}}}
\newcommand{\ey}{\bm{\hat{y}}}
\newcommand{\ez}{\bm{\hat{z}}}
\newcommand{\er}{\bm{\hat{r}}}
\newcommand{\etheta}{\bm{\hat{\theta}}}
\newcommand{\ek}{\bm{\hat{z}}}

\newcommand{\vHS}{\vit_{{\mathrm{shear}}}}

\newcolumntype{d}[1]{D{.}{.}{#1}}

\title{Zonal flow in a resonant precessing  cylinder} 
\shorttitle{Zonal flow in a resonant precessing  cylinder} 

\shortauthor{D. Gao, P. Meunier, S. Le Diz\`es and C. Eloy} 
\author{
	Donglai Gao\aff{1,2},
	Patrice Meunier\aff{1},
	St\'ephane Le Diz\`es\aff{1},
	and 
	Christophe Eloy\aff{1}\corresp{\email{celoy@ec-m.fr}}
}

\affiliation{
	\aff{1} Aix Marseille Univ, CNRS, Centrale Marseille, IRPHE, Marseille, France\\
	\aff{2} Joint laboratory of wind tunnel and wave flume,  Harbin Institute of Technology,  Harbin, P.~R. China
}

\begin{document}
	
	\maketitle
	
	\begin{abstract} 
A cylinder undergoes precession when it rotates around its axis and this axis itself rotates around another direction. 
In a precessing cylinder full of fluid, a steady and axisymmetric component of the azimuthal flow is generally present. This component is called a zonal flow.  
Although zonal flows have  been often observed in experiments and numerical simulations, their origin has eluded theoretical approaches so far. 
Here, we develop an asymptotic analysis to calculate the zonal flow forced in a resonant precessing cylinder, that is when the harmonic response is dominated by a single Kelvin mode. 
We find that the zonal flow originates from three different sources: 
(1) the nonlinear interaction of the inviscid Kelvin mode with its viscous correction; 
(2) the steady and axisymmetric response to the nonlinear interaction of the Kelvin mode with itself; and
(3) the nonlinear interactions in the end boundary layers. 
In a precessing cylinder, two additional sources arise due to the equatorial Coriolis force and the forced shear flow. However, they cancel exactly. The study thus generalises to any Kelvin mode, forced by precession or any other mechanism.  
The present theoretical predictions of the zonal flow are confirmed by comparison with numerical simulations and experimental results. 
We also show numerically that the zonal flow is always retrograde 
in a resonant precessing cylinder ($m=1$) or when it results from resonant Kelvin modes of azimuthal wavenumbers $m=2$, $3$, and presumably higher. 
\end{abstract}
	

\section{Introduction}
	
The zonal flow denotes the steady axisymmetric flow that is generated by the nonlinear interactions of unsteady motions in a rotating fluid. 
Its structure strongly depends on the geometry, the boundary conditions
and the forcing. In the present study, we provide an analytic expression for the zonal flow obtained in a precessing cylinder at resonance, that is when the harmonic response is dominated by a single 
Kelvin mode.  
	
The flow in a precessing cylinder has been examined  in an engineering context for its interesting mixing properties \citep{Meunier2020} and for its importance in the
stability of gyroscopes \citep{Stewartson1959,Gans1984,lambelin2009non}.
But most works have been motivated by geophysical and astrophysical applications.
Planets generally rotate and interact  gravitationally with neighboring stars, planets or satellites. 
These interactions may lead to periodic variations of the shape,  to 
changes in the direction of the rotation axis,  or to oscillations of the planet's rotation rate. 
They correspond to harmonic forcing with three different azimuthal wavenumbers: tide ($m=2$), precession, nutation, and latitudinal libration ($m=1$), and longitudinal libration ($m=0$), respectively. 
As recently reviewed in \cite{LeBars2015}, these forcings can drive important flows in the liquid core of planets. The question whether they can drive a dynamo has been 
the subject of many studies  \citep[e.g.,][]{Malkus1968,Tilgner2005,Wu2009,Wu2013,Cebron2014}. 
	
The flow in a precessing spheroid has been first described by the inviscid solution of \cite{Sloudsky1895} and  \cite{poincare1910}. 
	\cite{busse1968steady}  then considered  the viscous torque generated by the boundary layers to predict the slow down of the solid body rotation \citep[see also][]{Hollerbach1995,Kerswell1995}. This non-linear theory, which has been validated experimentally \citep{Noir2001,Horimoto2018,Nobili2021}, predicts a hysteresis cycle between two solutions for strong ellipticity or large tilt angles. In these studies, the zonal flow plays a  crucial role.

In parallel to these studies on spherical geometries akin to planets or satellites, considerable efforts have been devoted to the more academic case of a precessing cylinder. 
The early experiment of \cite{McEwan1970} modelled the precessional forcing by a rotating tilted top. \cite{McEwan1970} showed that the flow becomes resonant when the forcing frequency is equal to the frequency of an inertial eigenmode, known as a Kelvin mode \citep{Kelvin1880}. 
This resonance leads to a flow much larger than in the spheroidal geometry. 
This was later observed experimentally for precessing cylinders  \citep[e.g.,][]{manasseh1992,manasseh1994,manasseh1996,kobine1995inertial,Kobine1996}. In these early experiments, the flow was characterised mainly based on direct visualisations, measurements of torque, energy dissipation rate, and point-wise velocity.	

At moderate Ekman numbers, \cite{Gans1970} showed experimentally and theoretically that the amplitude of the resonant Kelvin mode is saturated by viscosity. Indeed, Ekman pumping damps the Kelvin mode, leading to a maximal resonant amplitude proportional to the tilt angle divided by the square root of the Ekman number. However, at small Ekman numbers, viscous effects can become weaker than nonlinear effects and \cite{Meunier2008} showed that the maximal amplitude is then proportional to the tilt angle to the power $\frac{1}{3}$. 

At resonance, \cite{Kobine1996} and \cite{Meunier2008} observed experimentally  that a strong zonal flow is generated by the forced Kelvin mode.
This zonal flow induces a detuning of the resonance, which saturates the amplitude of the Kelvin mode. 
It was also argued that the zonal flow can destabilise the base flow through a centrifugal instability \citep{Kobine1996,Giesecke2018} or a shear instability \citep{jiang2015}.	
To better understand the mechanism of saturation and these potential instabilities, it is thus important to determine the zonal flow produced. 

In both the tilted top experiments and the precessing experiments, the flow can become unstable for tilt angles as small as few degrees.
\cite{McEwan1970} proposed that a forced Kelvin mode can trigger a triadic resonance with two free Kelvin modes, leading to a parametric instability. \cite{Mason2002} investigated theoretically this instability inside a plane fluid layer in the limit of weak precession. They found that the triadic resonances indeed generate two free modes. Moreover, the triggered modes can also be unstable and further result in a secondary instability and then a tertiary instability and so on \citep{kerswell1999}. The onset  of precessional turbulence, its `breakdown', can thus be interpreted in the light of interactions between Kelvin modes. 

\cite{Lagrange2011,lagrange2016triadic} have confirmed this triadic resonance scenario through experiments in precessing cylinders. 
To predict the onset of triadic resonances, \cite{Lagrange2011,lagrange2016triadic} also proposed  a weakly nonlinear theory, which includes viscous effects and a heuristic model of the slow growth of a zonal flow due to the unstable modes. 
This latter effect leads to a detuning of the resonant mode, which in turn damps the triadic resonance instability \cite[see also][]{Herault2019}, and eventually cause intermittent cycles of growth and decay of the unstable flow. 
Numerical simulations recently confirmed this weakly nonlinear dynamics  \citep{Albrecht2015,Albrecht2018,Marques2015,Lopez2018} and have emphasised the central role played by the zonal flow in this dynamics. 
	 
In geophysical fluid dynamics, a zonal flow is defined as an axisymmetric azimuthal velocity. This zonal flow has also been called mean streaming flow \citep{Albrecht2020} since it is mostly generated by streaming through the action of Reynolds stresses \citep{riley2001steady}. It is important to note that this zonal flow may not be invariant along the axial direction. For example, \cite{Waleffe} showed that a Kelvin mode of amplitude $A$ generates a nonlinear flow at order $|A|^2$ with an axial wavenumber twice as large as the Kelvin mode wavenumber \citep[see also][]{Meunier2008}.

The particular case of a zonal flow invariant along the axis is called a geostrophic flow, because it is a solution of the geostrophic balance between the Coriolis force and the pressure gradient. \cite{Greenspan69} proved mathematically that a geostrophic flow cannot be generated by a nonlinear interaction of an inertial mode of amplitude $A$ with itself in the limit of small Ekman and Rossby numbers. However, \cite{Meunier2008} showed that a geostrophic flow can be weakly forced by the Ekman boundary layers at order $|A|^2 \Ek^{1/2}$ (where $\Ek$ is the Ekman number based on the radius and the angular velocity of the cylinder). 
This flow is saturated by viscous damping when its amplitude becomes an order larger in $\Ek^{-1/2}$ than the forcing, i.e. for an amplitude proportional to $|A|^2$ similar to a classical streaming flow. This mechanism makes the prediction of the geostrophic flow quite difficult, because it implies many sources of forcing at an order $\Ek^{1/2}$ smaller than the resulting azimuthal flow.
	
	In this paper, we calculate the zonal flow forced by a Kelvin mode, by focusing on a resonant precessing fluid cylinder at small Ekman numbers and weak precession. We first derive the flow forced by precession in \S\ref{Sec.FormulationOfTheProblem} and introduce the properties of the geostrophic flow in \S\ref{Sec.GeostrophicFlow}. We then calculate, with an asymptotic approach, the five components of the zonal flow in \S\ref{Sec.ForcingGeostrophic}. 
	 The proposed theoretical solution is then compared to the numerical results of \cite{Albrecht2020} in \S\ref{Sec.ComparisonNumerics} and the experimental results of \cite{Meunier2008} in \S\ref{Sec.ComparisonExperiments}. The sign of the angular momentum of the zonal flow (prograde or retrograde) is discussed in \S\ref{Sec.RetrogradeDirection}. Finally, some conclusions are drawn in \S\ref{Sec.Conclusion}.
	
\section{Precessing cylinder}
\label{Sec.FormulationOfTheProblem} 
	
\begin{figure} 
		\centerline{\includegraphics[width=0.5 \textwidth] {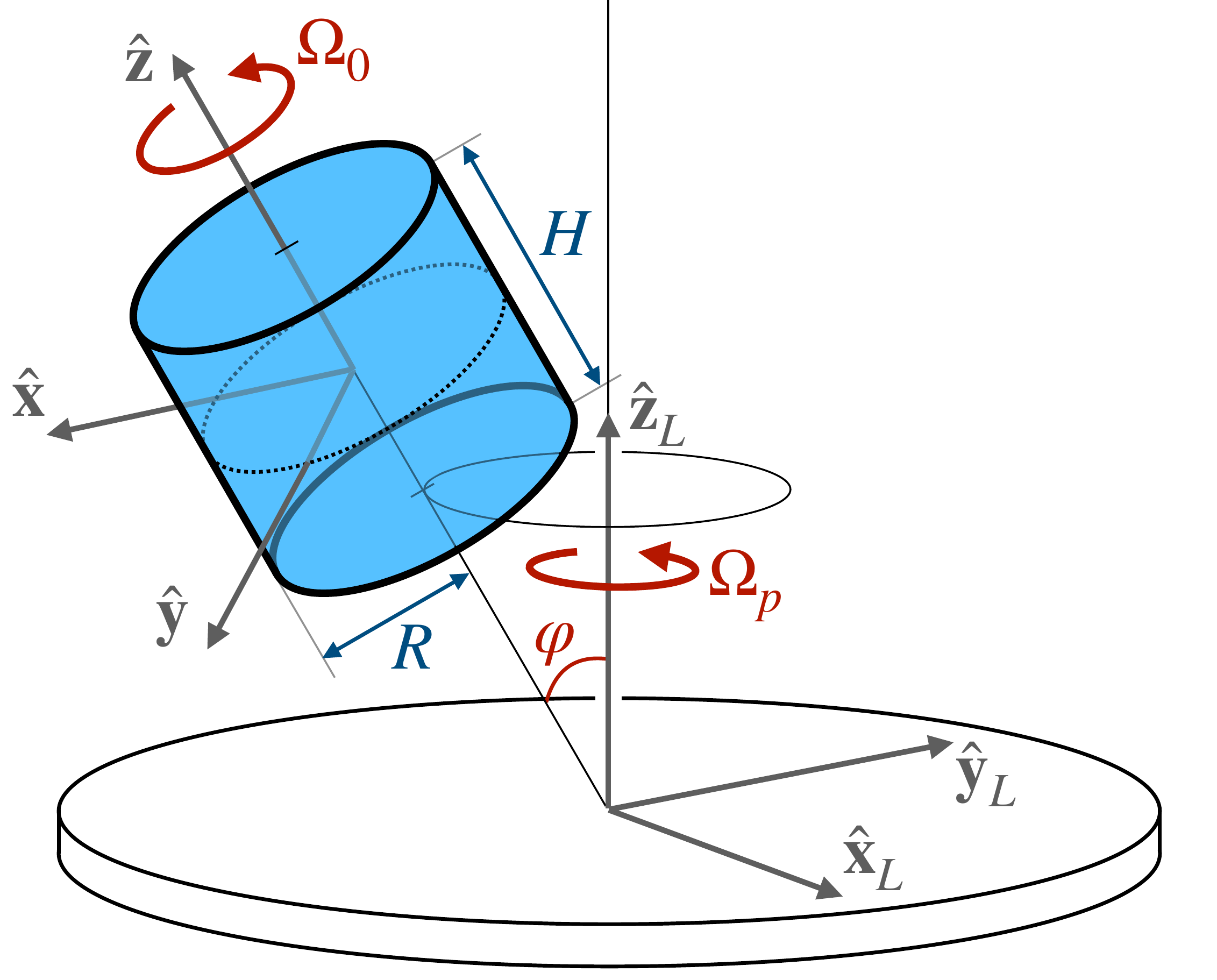}}
		\caption{Schematic representation of a fluid-filled precessing cylinder, with radius $R$ and height $H$. The cylinder rotates at  angular velocity $\Omega_0$ around $\ek$, which itself also precesses at angular velocity $\Omega_p$ around the vertical axis $\ek_L$ in the laboratory frame. The precession angle between these two axes is $\varphi$.}
		\label{fig1}
	\end{figure}

	Consider a cylinder of radius $R$ and height $H$, whose axis is along $\ek$ (Figure~\ref{fig1}). This cylinder is entirely filled with a Newtonian fluid of density $\rho$ and kinematic viscosity $\nu$. The cylinder rotates at  angular velocity $\Omega_0$ around its axis $\ek$, which also precesses at angular velocity $\Omega_p$ around the vertical axis $\ek_L$. We denote by $\varphi$ the tilt angle, i.e. the angle between $\ek$ and  $\ek_L$.
		
	To make the problem dimensionless, we use $R$, $\rho$ and $\Omega = \Omega_0+\Omega_p\cos\varphi$ as characteristic dimensions of length, density and frequency. The problem is associated to four dimensionless numbers: 
(1) the aspect ratio, $h=H/R$;
(2) the forcing frequency, $\omega =\Omega_0/\Omega$;
(3) the forcing amplitude, $\Ron=\Omega_p\sin\varphi/\Omega$;  
and (4) the Ekman number, $\Ek = \nu/\Omega^2 R$. 	
Dimensionless quantities will now be noted with lowercase letters.
The dimensionless flow velocity in the cylinder framework $(O,\,\ex,\,\ey,\,\ez)$ is denoted by $\vit={\bf{U}}/\left(R\Omega\right)$. We will use the cylindrical coordinates $(r,\theta,z)$, where $z=0$ corresponds to the mid-height  of the cylinder and  the position vector will be noted $\bf{r}$. 
	
	In the cylinder framework, the dimensionless Navier--Stokes equations are \citep{Meunier2008,Lagrange2011}
\begin{subequations}\label{EqNavierStokes}
\begin{equation}\label{EqNavier}
		\frac{\partial\vit}{\partial t} + 
		2 \ek\times \vit + \Nab p =  
		- 2\Ron\,\omega r\, \cos(\omega t+\theta)\,\ek
		- 2\Ron\,\del \times \vit
		 - \vit \cdot \Nab \vit + \Ek\,\Nab^2\vit,
\end{equation}
\begin{equation}
		\Nab\cdot\vit = 0,
\end{equation}
\end{subequations}
with $\del = \cos(\omega t)\,\ex -\sin(\omega t)\,\ey$. On the left hand side of \eqref{EqNavier}, the first term is inertia, the second term is the Coriolis force and $p$ is a dimensionless pressure field including all potential terms \citep{Meunier2008}. On the right hand side of \eqref{EqNavier}, the first term is the forcing due to precession, the second term is the equatorial Coriolis force, the third term is the convective nonlinear term, and the last term is the viscous force. 
	
We will now  consider the asymptotic limits of 
small Ekman number $\Ek$ and small forcing $\Ron$, which is achieved when the tilt angle is small or the Poincar\'e number $\Omega_p/\Omega_0$ is small.
We will seek a solution of the Navier--Stokes equations expressed as a series in powers of these small quantities $\Ek$  and $\Ron$.

\subsection{Inviscid solution}
	The base flow $\vitbase$ forced by  precession can be found by solving  the Navier-Stokes equations (\ref{EqNavierStokes}\textit{a},\textit{b}) at first order in $\Ron$. Following \cite{lagrange2016triadic}, we  only keep  the first term on the right hand side of \eqref{EqNavierStokes}. The solution $\vitbase$ is then composed of two parts: a particular solution in the form of a horizontal shear flow and a sum of Kelvin modes of azimuthal wavenumber $m=1$, which are the solutions of the homogeneous equation. Thus	
	\begin{equation}\label{EqHShear}
		\vitbase =
	\Ron \left( \vHS  + \sum\limits_{j = 1}^\infty A_j \qvit _j \right) \rme^{\rmi\left(\omega t + m\theta\right)} + \cc,
		\end{equation}
with $\cc$ meaning `complex conjugate'. 
In \eqref{EqHShear}, $\vHS  = \omega z(\rmi \er -\etheta)/(2 - \omega)$ is the horizontal shear, $A_j$ are the amplitudes of Kelvin modes of axial wavenumber $k_j = (2j-1)\pi /h$, and $\qvit _j$  their associated velocity fields given by
\begin{equation}\label{EqModeForce}
	\qvit _j = \left( \begin{array}{*{20}{r}}
	\rmi\,u \, \sin \left( k_j z \right) \\
	v \, \sin \left( k_j z \right) \\
	\rmi\,w \, \cos \left( k_j z \right) \\
	\end{array} \right), 
\end{equation}
with $u$, $v$ and $w$, real functions of $r$
\begin{subequations}\label{uvw}
		\begin{align}
		u(r) & = \frac{ - 1}{4 - \omega ^2 }\left( \omega \delta J'_m \left(\delta r\right) + \frac{2m}{r}J_m \left( \delta r \right) \right),\\
		v(r) & = \frac{1}{4 - \omega^2}\left( 2\delta J'_m \left( \delta r \right) + \frac{\omega m}{r}J_m \left( \delta r \right) \right), \\
		w(r) & = \frac{k}{\omega}J_m \left(\delta r\right),
		\end{align}
\end{subequations}
$J_m$ the Bessel function of the first kind,
$\delta = k_j(4/\omega^2 -1)^{1/2} $ playing the role of a radial wavenumber, and primes noting differentiation.
	
Here, we will assume that the forcing is resonant. This occurs when the forcing frequency takes particular values, $\omega=\omega_{n,j}$, such that $u(1)=0$ for a particular Kelvin mode. In that case, this Kelvin mode amplitude $A_j$ is asymptotically large and the radial wavenumber takes the value $\delta = \delta_{n,j} = k_j(4/\omega_{n,j}^2 -1)^{1/2}$. 	
Each resonance is associated to a couple of strictly positive integers $(n,j)$: 
$n$ numbers the radial wavenumber $\delta$ (the function $u(r)$ has exactly $n$ zeros for $0<r\le 1$), and
$j$ numbers the axial wavenumber $k_j$ such that $k_j = (2j-1)\pi /h$. In \cite{Meunier2008}, we also referred to these resonances by the phrase ``$j$-th resonance of the $n$-th Kelvin mode''.
	
At resonance, the amplitude $A_j$ of the resonant Kelvin mode can be calculated by either invoking viscous effects, in which case $A_j=O(\Ek^{-1/2})$, or  by invoking nonlinear effects, in which case $A_j=O(\Ron^{-2/3})$ \citep{Meunier2008}. 
In either case, the base flow is dominated by this single resonant Kelvin mode of amplitude $A=A_j$
\begin{equation}\label{EqSolLinH2} 
	\vitbase \approx
	A\, \qvit \, \rme^{\rmi\left(\omega t + m \theta\right)} + \cc,
\end{equation}
whose velocity field $\qvit$ is given by \eqref{EqModeForce} and we have dropped the indices $j$ for simplicity.

\subsection{Viscous solution}
	The base flow given in \eqref{EqSolLinH2} is a solution of the inviscid problem. To satisfy the viscous boundary condition on the walls, it has to be complemented by a boundary-layer solution confined in the wall regions of thickness $O(\Ek^{1/2})$. The total flow at leading order is then
\begin{equation}\label{eq.utot0}
	\vit^{(0)}_\mathrm{tot.} = A\left(\qvit + \tilde{\qvit} \right)\rme^{\rmi\left(\omega t + m \theta\right)} +\cc. 
\end{equation}
The viscous solution $\tilde{\qvit}$ can be expressed as a series in powers of $\Ek^{1/2}$. This viscous solution has a non-zero axial component $\tilde w$ on the  end walls at order $\Ek^{1/2}$, forced through Ekman pumping \citep{Meunier2008}
\begin{equation}
\tilde w (z=-h/2) = \Ek^{1/2} k^2 \frac{\rmi-1}{2\sqrt{2}\,\omega^2}
\left(
\frac{2+\omega }{(2 - \omega)^{1/2}} +
\rmi\frac{2-\omega}{(2 + \omega)^{1/2}}
\right)
J_m (\delta r) \sin(kh/2),
\end{equation}
with an opposite flow on the upper wall in $z=h/2$.

To balance these non-zero axial flows on the end walls, the inviscid solution has to be corrected at order $\Ek^{1/2}$. A simple way to achieve this correction is to write the axial wavenumber as a series in powers of $\Ek^{1/2}$ 
\begin{equation}\label{eq.k}
k = k^{(0)} + \Ek^{1/2} k^{(1)} + \cdots, \quad
\mbox{with } k^{(0)} = (2j-1)\pi/h, 
\end{equation}
and 
\begin{equation}\label{eq.k1}
k^{(1)} =  \frac{k^{(0)}}{\sqrt{2}\,\omega h}  (1 + \rmi)
\left(
\frac{2+\omega }{(2 - \omega)^{1/2}} +
\rmi\frac{2-\omega}{(2 + \omega)^{1/2}}
\right)
.
\end{equation}
A similar correction exists for the radial wavenumber $\delta$ since it is proportional to $k$. 
When there is no ambiguity,
 we will refer to $k^{(0)}$ as $k$ in the following.

\section{Geostrophic flow}
\label{Sec.GeostrophicFlow}
	The  geostrophic flow $\qvit_0$ is an axisymmetric and steady solution of the linearised Navier--Stokes equation. It thus satisfies the homogeneous equations
\begin{equation}\label{NSgeo}
		2\,\ek\times \qvit_0 + \Nab p_0 =  0 ,\quad 
		\Nab\cdot\qvit_0 = 0,
\end{equation}
whose solution is
\begin{equation}
	\qvit_0 = A_0 v_0(r) \,\etheta,
\end{equation}
where $v_0(r)$ can be any function satisfying $v_0(0)=0$. Note that the general solution could also include an axial flow $w_0(r)$ independent of $z$, which is zero in our case because of the boundary conditions on the end walls. 
	
	Similarly to Kelvin modes, this solution of the inviscid problem has to be complemented by a boundary-layer solution in the regions close to the end walls. In these regions of thickness $O(\Ek^{1/2})$, the total flow is given by
	\begin{equation}\label{eqBLvo}
	\vit_0 = \qvit_0(r,z) + \tilde{\qvit}_0(r,\tilde{z}) +\cc,\quad \mbox{with }  
	\tilde{z} = \Ek^{-1/2} \left( z + \frac{h}{2} \right),
	\end{equation}
	for the lower wall. 
	
	Using this rescaled variable $\tilde{z}$ and solving the linearised Navier--Stokes equation with the proper boundary conditions ($\tilde{\qvit}_0 = - \qvit_0(z=-h/2)$ for $\tilde{z} =0$ and $\tilde{\qvit}_0 \to 0$ for $\tilde{z}\to\infty$), one can write the solution as a series in powers of $\Ek^{1/2}$
	\begin{equation}
	\tilde{\qvit}_0 =  \tilde{\qvit}_0^{(0)} + \Ek^{1/2}  \tilde{\qvit}_0^{(1)} + \cdots,
	\end{equation}
	the leading order solution is 
	\begin{equation}
	\tilde{\qvit}_0^{(0)} (r,\tilde{z}) = -\frac{A_0}{2} v_0(r)  \left( \begin{array}{c}
	\rmi\rme^{-(1+\rmi)\tilde{z}} - \rmi\rme^{-(1-\rmi)\tilde{z}} \\
	\rme^{-(1+\rmi)\tilde{z}} + \rme^{-(1-\rmi)\tilde{z}} \\
	0\\
	\end{array} \right).
	\end{equation}	
At next order, an axial flow is forced by Ekman pumping
	\begin{equation}
	\tilde{w}_0^{(1)}(r,\tilde{z}) = - \frac{A_0}{2r}  \frac{\rmd\left( r v_0 \right)}{\rmd r} 
	\left( \frac{1}{(1-\rmi)}\rme^{-(1+\rmi)\tilde{z}} + \frac{1}{(1+\rmi)}\rme^{-(1-\rmi)\tilde{z}}  \right),
	\end{equation}
	which means that there is an inflow of order $ O(\Ek^{1/2}v_0)$ on the lower wall of velocity
	\begin{equation}\label{eqw0v0}
	\tilde{w}_0 (z = -h/2) = - \Ek^{1/2}\frac{A_0}{2r} \frac{\rmd\left( r v_0 \right)}{\rmd r},
	\end{equation}  
	and, by symmetry, there is the opposite flow on the upper wall. 
	
The important point here is that the inviscid geostrophic flow $\qvit_0$ is  not directly forced, since \eqref{NSgeo} is homogeneous. But the geostrophic mode can be indirectly forced by symmetric inflows on the cylinder end walls (i.e. $w(z=h/2) = -w(z=-h/2)$). As we shall see in the next section, non-zero symmetric inflows can be created by the precession and the forced Kelvin mode. 
To cancel these inflows,  opposite inflows of the form $\tilde{w}_0$ are needed such that $\tilde{w}_0 (z = -h/2) = -w(z=-h/2)$. 
It means that a geostrophic flow $\qvit_0$ can be forced at an order $\Ek^{-1/2}$ higher than the non-zero inflow $w(z=h/2)$ through \eqref{eqw0v0}. 

We will now see how non-zero symmetric inflows can be created in precessing flows by four different mechanisms. 
	
	~
	
\section{Forcing of the geostrophic flow} \label{Sec.ForcingGeostrophic}
\subsection{Interaction between equatorial Coriolis force and the Kelvin mode}\label{sec5}
	
The first type of flow that forces a geostrophic flow is $\qvit_k$, the axisymmetric and steady solution of the linearised Navier--Stokes equation forced by the equatorial Coriolis force, i.e. the second term on the right-hand side of \eqref{EqNavierStokes}. It thus satisfies the following equations
\begin{equation}\label{eq.NSvk}
		2\,\ek\times \qvit_{k} + \Nab p_{k} = \mathbf{F}_{k},  \quad
		\Nab\cdot\qvit_{k} = 0,
\quad\mbox{with }
		\mathbf{F}_{k} = \Ron A \,\dd \times \qvit,
\end{equation}
and $\dd = -\er + \rmi \,\etheta$.
A particular solution of this forced equation gives an axial component of velocity 
	\begin{equation}\label{eq.vk}
		w_k = -\rmi\, \Ron  A \, \frac{1}{2kr}\frac{\rmd(r w)}{\rmd r} \sin (kz).
	\end{equation}
	In order to cancel the resulting inflow on the end walls, a geostrophic flow is necessary such that its viscous part satisfies $\tilde{w}_0 = -w_k$ on the lower wall. Using \eqref{eqw0v0}, it then leads to a geostrophic flow of order $\Ek^{-1/2}$ higher than $\qvit_k$ itself, with azimuthal velocity
\begin{equation}\label{eq.v0k}
\left(\vit_0\right)_{k} = -\Ron   \Ek^{-1/2} \Imag(A)   \frac{2 w}{k} \sin (kh/2) \, \etheta,
\end{equation}
with $\sin (kh/2) = (-1)^{j-1}$, because $k=(2j-1)\pi/h$. Note that velocities noted with the bold letter `u', like $\left(\vit_0\right)_{k}$, refer to the real component of the flow and include the complex conjugate, contrarily to velocities denoted by a bold `v', like $\qvit_{k}$ for instance. 

\subsection{Interaction between the Kelvin mode and the shear flow}
The second flow that forces a geostrophic flow is $\qvit_\mathrm{shear}$, the flow forced by the nonlinear interaction of the Kelvin mode with the shear flow. This flow is a solution of 
\begin{equation}\label{eq.NSvshear}
	2\,\ek\times \qvit_\mathrm{shear} + \Nab p_\mathrm{shear} =  \mathbf{F}_\mathrm{shear}, \quad	
	\Nab\cdot\qvit_\mathrm{shear} = 0,
\end{equation}
with
\begin{equation}\label{eq:Fshear}
\mathbf{F}_\mathrm{shear} = -\Ron (A\,\overline{\vit}_\mathrm{shear} \cdot \Nab  \qvit + \bar{A}\, \overline{\qvit} \cdot \Nab {\vit}_\mathrm{shear}).
\end{equation}

The component $\mathbf{F}_\mathrm{shear} \cdot \etheta$ has a non-zero real part, which yields through the incompressibility condition to a non-zero axial component of $\qvit_\mathrm{shear}$ 
\begin{equation}
w_\mathrm{shear} = \Ron \, \Imag(A) \, \frac{1}{2k r}\frac{\partial(r q_\mathrm{shear})}{\partial r},		
\end{equation}
with
\begin{equation}
q_\mathrm{shear} = - w(r) \sin(kz) + w(r) \frac{2 \delta \omega z}{(2-\omega)\sqrt{4-\omega^2}} \cos(kz) .  
\end{equation}
At the end walls $z=\pm h/2$, this axial flow balances exactly the axial flow $w_k$ given by \eqref{eq.vk}. It means that a geostrophic flow is created that cancels $\left(\vit_0\right)_{k}$
\begin{equation}\label{eq.v0shear}
\left(\vit_0\right)_\mathrm{shear} = -\left(\vit_0\right)_{k},
\end{equation}
with $\left(\vit_0\right)_{k}$ given by \eqref{eq.v0k}.
	
\subsection{Interaction of the Kelvin mode with itself in the bulk}
	\label{sec4}
	The third flow that forces a geostrophic flow is the mode $\qvit_{2k}$, the axisymmetric and steady solution of the linearised Navier--Stokes equation forced by the nonlinear interactions of the Kelvin mode with itself, which satisfies 	\begin{equation}\label{eq.NSv2k}
	2\,\ek\times \qvit_{2k} + \Nab p_{2k} =  \mathbf{F}_{2k}, \quad	
	\Nab\cdot\qvit_{2k} = 0,
	\quad\mbox{with }
	\mathbf{F}_{2k} = -|A|^2 \,
	\overline{\qvit} \cdot \Nab  \qvit.
\end{equation}
Note that the radial and axial components of the forcing are much larger than the azimuthal component, which vanishes for an inviscid Kelvin mode.
The azimuthal component $\mathbf{F}_{2k} \cdot \etheta$ is $O( |A|^2 \Ek^{1/2} )$ because of the viscous correction of the Kelvin mode given in \eqref{eq.k}. 
This azimuthal component of the forcing is treated below in \S\ref{sec.v0_Ek} through the component $\left(\vit_0\right)_\mathit{Ek}$ of the geostrophic flow. We will first examine the geostrophic flow due to the radial and axial forcing.
	
\subsubsection{Forcing at order $|A|^2$}	
	
At leading order $|A|^2$, a particular solution of \eqref{eq.NSv2k} has been given by \cite{Waleffe}
	\begin{equation} \label{eq.v2k}
	\qvit_{2k} = - |A|^2 v_{2k}(r) \cos(kh) \cos\left( 2 k z\right)\etheta, \quad \mbox{with }
	v_{2k}(r) = \frac{k}{\omega} \left(\frac{1}{2k}\frac{\rmd (uv)}{\rmd r} -  vw \right),
	 \end{equation}
with $u$, $v$ and $w$ the radial functions of the Kelvin mode, given in \eqref{uvw}.

	The velocity $\qvit_{2k}$ is solution of the inviscid problem. Although this solution seems to violate the statement of \cite{Greenspan69} that no geostrophic flow can be generated by an inviscid inertial mode,  in fact it does not, since $\qvit_{2k}$ is a non-geostrophic zonal flow (because of its dependence  on $z$). The associated boundary layer solution can be found by the same method as the one described above for the geostrophic flow in \S\ref{Sec.GeostrophicFlow}. It yields an inflow on the lower wall that can be written
	\begin{equation}
	\tilde{w}_{2k}( z=-h/2) = |A|^2\Ek^{1/2} \frac{1}{2r} \frac{\rmd\left( r v_{2k} \right)}{\rmd r}.
	\end{equation}  
	
	To cancel this inflow, a geostrophic flow is necessary that will produce the opposite flow. This geostrophic flow is found using \eqref{eqw0v0} and can be written
\begin{equation}\label{eq.v02k}
\left(\vit_0\right)_{2k} = 2\,|A|^2  v_{2k}(r)\,  \etheta,
\end{equation}	
	Note that the sum of the flow $\qvit_{2k}$ and the above geostrophic flow gives a real component of the velocity
	\begin{equation} \label{eq.v2k+v02k}
	\vit_{2k} + \left(\vit_0\right)_{2k}= 2\, |A|^2 v_{2k}(r) 
	\left(1 - \cos(kh) \cos\left( 2 k z\right) \right)  \, \etheta,
	\end{equation}
that cancels for $z=\pm h/2$. 

\subsubsection{Forcing at order $|A|^2 \Ek^{1/2}$}\label{sec.v0_Ek}	

At order $|A|^2 \Ek^{1/2}$, the inviscid Kelvin mode interacts nonlinearly with its viscous correction, which can be obtained by a series expansion of $k$ in powers of $\Ek^{1/2}$ given in \eqref{eq.k1}.  
This nonlinear interaction can also be viewed as the interaction between the Kelvin mode and the flow resulting from Ekman pumping at the end walls. It yields a non-zero azimuthal forcing $\mathbf{F}_{2k} \cdot \etheta = O( |A|^2 \Ek^{1/2} )$.

Through \eqref{eq.NSv2k}, this forcing gives rise to a radial velocity $u_\mathit{Ek} = \frac{1}{2}\mathbf{F}_{2k} \cdot \etheta$.
This radial component induces an axial flow $w_\mathit{Ek}$ through the incompressibility condition
\begin{equation}\label{eq.w2kbis}
w_\mathit{Ek} (z=-h/2) = |A|^2 \Ek^{1/2} \frac{1}{2r} \frac{\rmd\left( r v_\mathit{Ek} \right)}{\rmd r}, \quad
\mbox{with }
v_\mathit{Ek}(r) = \frac{h }{\omega} u w\,\Imag\left(k^{(1)}\right),
\end{equation}
where $k^{(1)}$ is the correction to the axial wavenumber given in \eqref{eq.k1} and $v_\mathit{Ek}$ is found by integrating $\mathbf{F}_{2k} \cdot \etheta$ between $z=-h/2$ and 0 (see also \eqref{eq.v0generic} below).

To cancel this flow, a geostrophic flow $\left(\vit_0\right)_\mathit{Ek}$ is necessary 
\begin{equation}\label{eq.v0Ek}
\left(\vit_0\right)_\mathit{Ek} = 2 |A|^2  v_\mathit{Ek}(r)\,  \etheta.
\end{equation}

\subsection{Interactions of the Kelvin mode with itself in the boundary layers}
\label{sec6}
Last, we consider the flow $\tilde{\qvit}_\mathit{BL} $ forced in the end wall boundary layers through nonlinear interactions of the viscous Kelvin mode with itself. Without loss of generality, we consider the lower wall boundary layer of thickness $O(\Ek^{1/2})$. In this boundary layer, at leading order, the total flow $\qvit^{(0)}_\mathrm{tot.}$ can be written as 
the sum of the inviscid solution and its viscous correction	\begin{equation}\label{eq.utotBL}
	\qvit^{(0)}_\mathrm{tot.} =
	A \sin \left(-kh/2\right) \left( \begin{array}{c}
	\rmi\, u_\mathrm{tot.}   \\
	v_\mathrm{tot.}  \\
	\Ek^{1/2} \rmi\, w_\mathrm{tot.}   \\
	\end{array} \right)  \rme^{\rmi\left(\omega t + m\theta\right)},
	\end{equation}
where $u_\mathrm{tot.}$, $v_\mathrm{tot.}$ and $w_\mathrm{tot.}$ are  functions of $r$ and $\tilde{z}$, the rescaled vertical coordinate given by \eqref{eqBLvo}. These functions are given by (\ref{eq.uvwtot}\textit{a}--\textit{c}) in Appendix~\ref{appA}. 
	
Through nonlinear interaction with itself, this flow acts as a steady and axisymmetric forcing of the linearised Navier--Stokes equations
	\begin{subequations}\label{eq.NS.BL}
		\begin{align}
		\Ek^{-1/2} \frac{\partial \tilde{p}_\mathit{BL}  }{\partial \tilde{z}} \ez +
		2\,\ek\times \tilde{\qvit}_\mathit{BL}  + 
		\frac{\partial \tilde{p}_\mathit{BL}  }{\partial r} \er - 
		\frac{\partial^2 \tilde{\qvit}_\mathit{BL} }{\partial \tilde{z}^2} & = 
		\mathbf{F}_\mathit{BL} , \\  
		\Ek^{-1/2} \frac{\partial \tilde{w}_\mathit{BL} }{\partial \tilde{z}} +
		\frac{1}{r} \frac{\partial\left( r \tilde{u}_\mathit{BL}  \right)}{\partial r} & = 0.
		\end{align}
	\end{subequations}
where the forcing term can be written as
	\begin{equation}\label{eq:NLforcing}
	\mathbf{F}_\mathit{BL} (r,\tilde{z}) = -{\overline\qvit}^{(0)}_\mathrm{tot.} \cdot \Nab \qvit^{(0)}_\mathrm{tot.} = 
	|A|^2 \sum_i \left( \begin{array}{c}
	a_i + c_i \tilde{z} \\
	\rmi b_i + \rmi d_i \tilde{z} \\
	O(\Ek^{1/2})
	\end{array} \right) \rme^{-\kappa_i \tilde{z}},
	\end{equation}
with the scalars $\kappa_i$ and the functions of $r$, $a_i$, $b_i$, $c_i$, and $d_i$, are given in Appendix~\ref{appA}. 
	
Equations (\ref{eq.NS.BL}\textit{a}--\textit{b}) can be solved by expanding $\tilde{\qvit}_\mathit{BL} $ in a series of powers of $\Ek^{1/2}$ and using the boundary conditions, $\tilde{\qvit}_\mathit{BL}  = 0$ for $\tilde{z} =0$ and $\tilde{z}\to\infty$. 
The solution at relevant order is given by \eqref{eq.vBL0sol} in Appendix~\ref{appA}.
This solution implies a non-zero axial flow $\tilde{w}_\mathit{BL} $ at order $|A|^2 \Ek^{1/2}$ on the lower end wall. 
To cancel this inflow, a geostrophic flow is necessary at order $|A|^2$ such that
\begin{equation}\label{eq.v0NL}
\left(\vit_0\right)_\mathit{BL}  = 2 |A|^2 v_\mathit{BL} (r)\, \etheta, \end{equation}
with
	\begin{equation}\label{eq.v0NLr}
	v_\mathit{BL} (r) = \Real\left[
	\sum_i 
	\frac{a_i \kappa_i+\rmi  b_i (\kappa_i+2))}{\kappa_i (\kappa_i (\kappa_i+2)+2)}
	+ 2 \frac{c_i (\kappa_i+1) \kappa_i^2+\rmi d_i (\kappa_i (\kappa_i+2)^2+2)}{\kappa_i^2 (\kappa_i (\kappa_i+2)+2)^2} \right] .
	\end{equation}
	
\subsection{Total geostrophic flow}

We now have calculated the 5 components of the geostrophic flows originating from 5 different inflows at the end walls. These components of the geostrophic flow, noted $\left(\vit_0\right)_{k}$, $\left(\vit_0\right)_\mathrm{shear}$, $\left(\vit_0\right)_{2k}$, $\left(\vit_0\right)_\mathit{Ek}$ and $\left(\vit_0\right)_\mathit{BL} $, are given respectively by \eqref{eq.v0k}, \eqref{eq.v0shear},\eqref{eq.v02k}, \eqref{eq.v0Ek} and \eqref{eq.v0NL}. 
Summing them up gives the total forced geostrophic flow 
\begin{equation}\label{eq.v0tot}
\vit_0 
	= \left(\vit_0\right)_k + \left(\vit_0\right)_\mathrm{shear} + \left(\vit_0\right)_{2k} + \left(\vit_0\right)_\mathit{Ek} + \left(\vit_0\right)_\mathit{BL}  
	= A_0 v_0(r) \,\etheta,
\end{equation}
with
\begin{equation}\label{eq.v0totr}
A_0 = |A|^2, \quad
v_0(r) = 2v_{2k}(r) + 2 v_\mathit{Ek}(r) + 2 v_\mathit{BL} (r),
\end{equation}
and $v_{2k}$, $v_\mathit{Ek}$ and $v_\mathit{BL} $ given by \eqref{eq.v2k}, \eqref{eq.w2kbis} and \eqref{eq.v0NLr}. 

Note that the components $\left(\vit_0\right)_k$ and  $\left(\vit_0\right)_\mathrm{shear}$ exactly cancel each other, such that the amplitude of the geostrophic flow is proportional to $|A|^2$ and does not include any terms proportional to $\Ron A \Ek^{-1/2}$. Because these two terms balance exactly, the resulting geostrophic flow does not depend on the precessional forcing anymore. The present results are thus applicable to any Kelvin mode, independently of the way it has been forced. In particular, as we shall see below, our calculations of the geostrophic flow is valid for Kelvin modes of arbitrary azimuthal wavenumber $m$.

\subsection{Weakly non-linear amplitude equations}
So far, we have calculated the amplitude of the geostrophic flow $A_0$ in the steady regime, and we have found $A_0=|A|^2$. Based on this calculation and 
following the weakly nonlinear calculation in \cite{Meunier2008}, we can easily obtain the weakly non-linear amplitude equations describing transient regimes. 

To obtain the dynamic equations for $A$ and $A_0$, we have to correct the term describing non-linear interaction between the geostrophic flow and the Kelvin mode. In \cite{Meunier2008}, we did not take into account the components $\left(\vit_0\right)_{2k}$ and $\left(\vit_0\right)_\mathit{Ek}$ of the geostrophic flow and there was also an error in the calculation of $\left(\vit_0\right)_\mathit{Ek}$ (in particular, the term $w_4 \tilde z$ appearing in \eqref{eq.uvwtotc} was missing). 
At resonance, the dynamic amplitude equations are then given by	
\begin{subequations}\label{eq:weaklyNLamplitudeEq}
\begin{align}	
\frac{\rmd A}{\rmd t} & = 
	\Ron \,\rmi f 
	- \left(\Ek^{1/2}\mu + \Ek\, \nu\right) A
	+ \rmi \left(\sigma |A|^2 + \xi A_0 \right) A,\\
\frac{\rmd A_0}{\rmd t} & = \Ek^{1/2}\frac{2}{h}\left(|A|^2 - A_0\right),
\end{align}	
\end{subequations}
where the first 4 coefficients, $f$, $\mu$, $\nu$ and $\sigma$, can be calculated from \cite{Meunier2008} and are given in Table~\ref{tab_NL} for different resonances. 
The coefficient $f$ corresponds to the forcing by precession while $\mu$ and $\nu$ correspond to the Ekman and volume damping. 
The coefficient $\sigma$ corresponds to the non-linear coupling of the Kelvin mode with the non-geostrophic flow $\qvit_{2k}$ and with an unsteady elliptic flow noted $\qvit_{2\omega}$ in \cite{Meunier2008}. 

\begin{figure} 
\centerline{\includegraphics[scale=0.75]{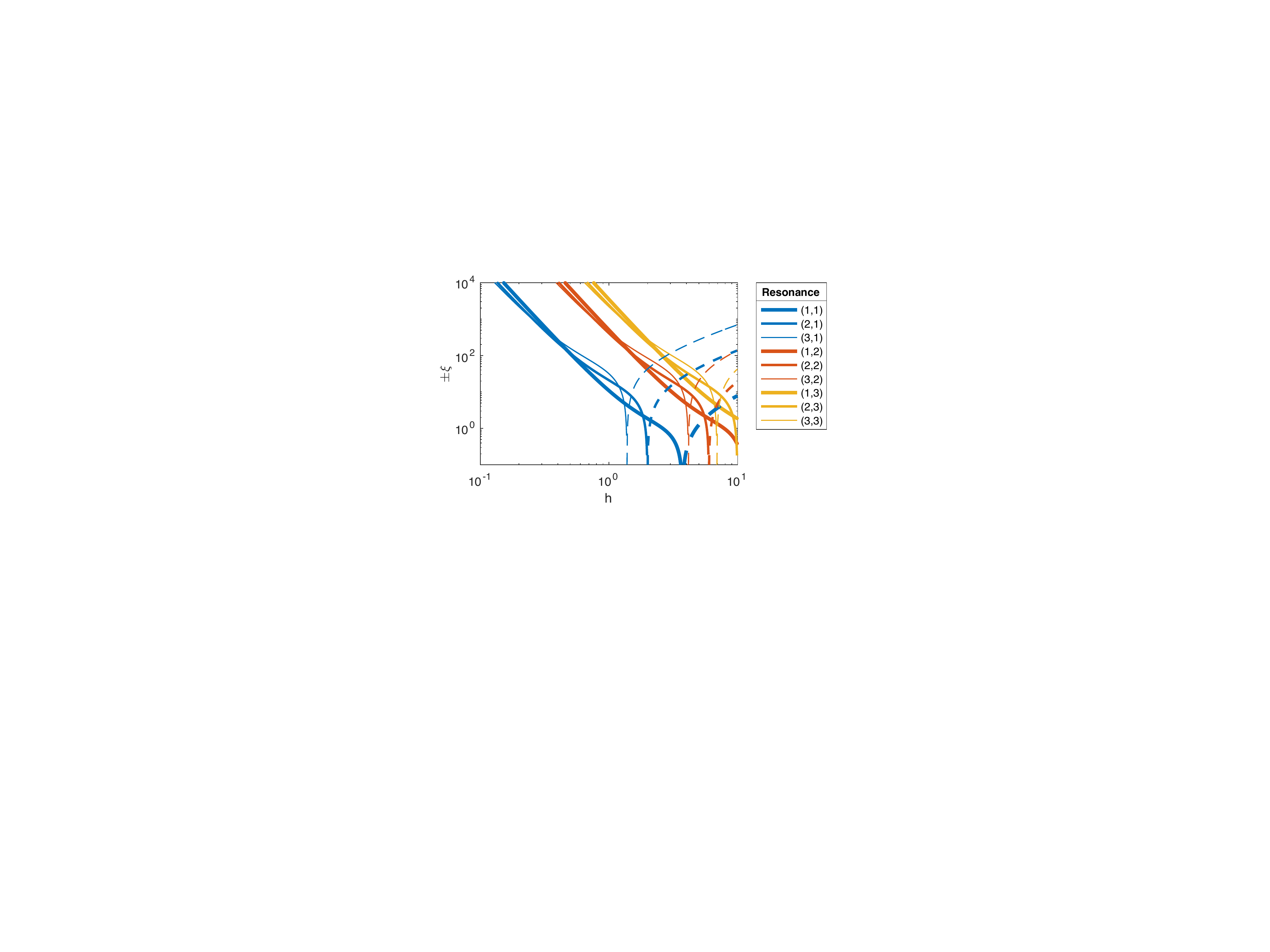}}
\caption{Value of the weakly nonlinear coefficient $\xi$ as a function of the aspect ratio $h$. The value is given for different resonances $(n,j)$ as noted in the legend and we use the convention of a solid line for $\xi$ negative and a dashed line for $\xi$ positive.}
\label{f.xi}
\end{figure}

The weakly nonlinear coefficient $\xi$, accounting for the nonlinear coupling between the Kelvin mode and the geostrophic flow, can be calculated as follows. We first calculate the nonlinear interaction of the Kelvin mode with the geostrophic flow: $\vit_0 \cdot \nabla \qvit +  \qvit\cdot \nabla \vit_0$.  We then project this forcing onto the Kelvin mode using the natural Hermitian product over the cylinder volume. The coefficient $\xi$ is simply this Hermitian product normalised by the norm of the Kelvin mode (which is nothing else than its kinetic energy). It can be written $\xi = X/E$ with 
\begin{equation}\label{eq.xi1}
	X = 2 \pi h |A|^2 \int_0^1 \left( 
	\frac{2k}{\omega} u w\, v_0 -  u v \frac{1}{r} \frac{\rmd (r v_0)}{\rmd r} 
	\right)r\, \rmd r,
\end{equation}
$E$ the kinetic energy of the mode
\begin{equation}\label{eq.E}
	E =  2 \pi h |A|^2 \frac{2 k^2 + m (2m - \omega)} {(4 - \omega^2)\, \omega^2}  J_m(\delta)^2,
\end{equation}
and $v_0(r)$ given by \eqref{eq.v0totr}. The values of $\xi$ for the first nine resonances are plotted in figure~\ref{f.xi}. It shows that $\xi$ scales as $-h^{-4}$ for $h\ll 1$, changes sign for $h=O(1)$ and scales as $h$ for $h\gg 1$. 
	
\begin{table}
\caption{Coefficients appearing in the weakly nonlinear amplitude equations \eqref{eq:weaklyNLamplitudeEq} for different resonances $(n,j)$ and different aspect ratios $h$.}
\begin{center}
\begin{tabular}{c*{3}{D..{-1}}@{\hskip 15pt}c@{\hskip 10pt}*{3}{D..{-1}}}
$(n,j)$  & $h$  & \multicolumn{1}{c}{$\omega_{n,j}$} & \multicolumn{1}{c}{$f$}	& $\mu$ 		& \nu 	& \sigma 	& \xi 	\\[3pt]
$(1,1)$  & 1.62	& 1.181	&  0.452 	& $1.86 - 0.42\rmi$	&  10.78	&   0.13	&   -2.94 \\
$(1,1)$  & 1.8	& 1.088	&  0.467 	& $1.80 - 0.27\rmi$	&  10.29	&  -0.06	&   -2.31 \\
$(1,1)$  & 2	& 0.996	&  0.469 	& $1.73 - 0.13\rmi$	&   9.96	&   0.23	&   -1.84 \\
$(2,1)$  & 2	& 0.510	& -0.074 	& $1.58 + 0.11\rmi$	&  37.99	& -10.69	&    0.14 \\
$(3,1)$  & 2	& 0.339	&  0.025 	& $1.47 + 0.16\rmi$	&  85.90	& -59.92	&   43.03 \\
$(1,2)$  & 2	& 1.774	& -0.042 	& $0.84 - 0.40\rmi$	&  28.24	&   7.80	&  -40.97 \\
$(2,2)$  & 2	& 1.285	&  0.021 	& $1.40 - 0.26\rmi$	&  53.77	&   9.96	&  -53.06 \\
$(3,2)$  & 2	& 0.946	& -0.009 	& $1.52 - 0.09\rmi$	&  99.28	&   2.54	&  -84.55 \\
$(1,3)$  & 2	& 1.911	& -0.007	& $0.50 - 0.32\rmi$	&  67.56	&  50.93	& -255.44 \\
$(2,3)$  & 2	& 1.632	& -0.006 	& $1.05 - 0.39\rmi$	&  92.60	&  66.39	& -225.67 \\
$(3,3)$  & 2	& 1.339	&  0.004 	& $1.34 - 0.28\rmi$	& 137.55	&  61.01	& -272.52 \\
\end{tabular}
\end{center}
\label{tab_NL}
\end{table}%

This weakly nonlinear analysis will be used below to predict the amplitude $A_0$ of the geostrophic flow when nonlinear effects are not negligible. 
	
\section{Comparison with numerical results}
\label{Sec.ComparisonNumerics}
		
We first examine the resonance  $(n=1,j=1)$, which occurs when the radial velocity $u(r)$ of the resonant  Kelvin mode has one zero for $0<r\le 1$ (this zero being in $r=1$) and when the wavenumber is $k=\pi/h$.  
	
In order to compare our theoretical results with the numerical results of \cite{Albrecht2020}, we focus on the aspect ratio $h=1.62$ for which the $(1,1)$ resonance is obtained at a forcing frequency $\omega_{1,1}=1.181$ (table~\ref{tab_NL}). The tilt angle  is  small, $\varphi=0.4^\circ$, and corresponds to a forcing amplitude $\Ron=-0.00127$, such that there is no instability in the numerical simulations at $\Ek=1.538 \times 10^{-4}$.

For these parameters, the forced Kelvin mode is mainly saturated by viscous effects and we can neglect nonlinear terms in (\ref{eq:weaklyNLamplitudeEq}\textit{a}). The Kelvin mode amplitude is then
 \begin{equation}
 A=\frac{\Ron \rmi f}{\Ek^{1/2}\mu + \Ek\, \nu} \label{e:AmplForcedMode}
 \end{equation}
where the coefficients $f$, $\mu$ and $\nu$, given in table~\ref{tab_NL}, correspond to $A=0.0047 -0.0221 \,\rmi$.
Using this value and \eqref{eq.E}, the total kinetic energy is found to be $E=0.0025$. This kinetic energy is approximately 30\% lower than the kinetic energy found in numerical simulations by isolating the azimuthal wavenumber $m=1$ \citep{Albrecht2020}. Equation \eqref{e:AmplForcedMode} thus underestimates the amplitude $A$ by about 14\% in this case.

\subsection{Spatial structure of the forcing}\label{subsec.compnum}

\begin{figure} 
\centerline{\includegraphics[scale=0.75]{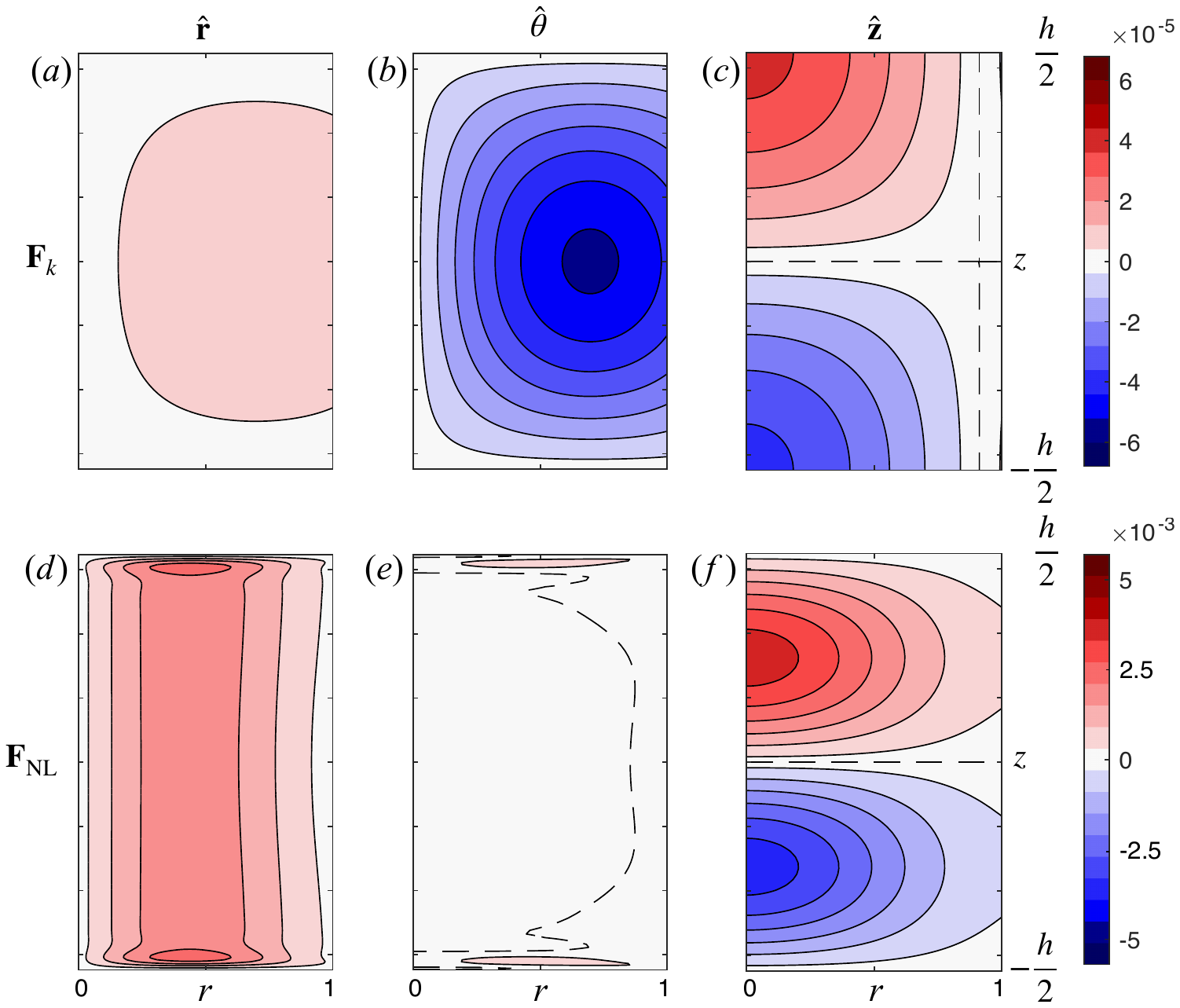}}
\caption{Forcing terms of the different components of the geostrophic flow. The columns are, from left to right, the radial (a, d), azimuthal (b, e), and axial (c, f) components of the forcing terms. 
The components of the forcing term $\mathbf{F}_k$ given in \eqref{eq.NSvk} are shown in a--c. 
The components of the forcing  term $\mathbf{F}_\mathrm{NL} = \mathbf{F}_\mathrm{shear} + \mathbf{F}_{2k} + \mathbf{F}_\mathit{BL} $ given in \eqref{eq:Fshear}, \eqref{eq.NSv2k} and \eqref{eq:NLforcing} are shown in d--f.  
The dashed line shows the level line for zero. 
The  parameters are $h=1.62$, $\Ek=1.5\times 10^{-4}$, $\omega=1.181$ and $\Ron=-0.00127$. To ease the comparison with figure~6 of \cite{Albrecht2020}, we use the same contours.}
\label{f.ForcingRes1}
\end{figure}

In the Navier--Stokes equations, the interaction of the forced Kelvin mode with the time-dependent precession forcing gives rise to a forcing term $\mathbf{F}_k$ given in \eqref{eq.NSvk}. This term is noted $\mathcal{C}^{\mathrm{inv}}$ in the  paper of \cite{Albrecht2020}. The three components of $\mathbf{F}_k$ along $\er$, $\etheta$ and $\ez$ are plotted in figure~\ref{f.ForcingRes1}\textit{a}--\textit{c}. 
They are in excellent agreement with the numerical results of \cite{Albrecht2020} (see their figures~6\textit{d}--\textit{f}). The largest component is the $\theta$-component, which has a unique negative lobe with a minimum value equal to $-5.4 \times 10^{-5}$, while in the numerical simulations this value is approximately $-6 \times 10^{-5}$ \citep{Albrecht2020}. A small difference between the numerics and the theory is also found for the axial component: the numerical results have a 20\% smaller amplitude; they exhibit small undulations of the iso-contours (probably due to inertial waves of small amplitudes emitted from the corners); and, in the simulations, this term vanishes at the top and the bottom because of the viscous boundary layers.

The nonlinear forcing terms can be gathered into a single forcing term $\mathbf{F}_\mathrm{NL} = \mathbf{F}_\mathrm{shear} + \mathbf{F}_{2k} + \mathbf{F}_\mathit{BL} $, which is noted  
 $\mathcal{R}^{\mathrm{inv}}$ in \cite{Albrecht2020}.
The three components of  $\mathbf{F}_\mathrm{NL}$ are plotted in figure~\ref{f.ForcingRes1}\textit{d}--\textit{f}. 
The forcing $\mathbf{F}_\mathrm{NL}$ gathers all nonlinear interactions of the forced flow with itself, including the Kelvin mode at order $A$, but also the particular shear flow solution at a lower order $O(\Ron)$, given by \eqref{EqHShear}, and the viscous correction to the Kelvin mode, also at a lower order $O(A\Ek^{1/2})$, obtained by a series expansion of the wavenumber $k$ given in \eqref{eq.k1}.
Figure~\ref{f.ForcingRes1}\textit{d}--\textit{f} shows an excellent  agreement with the numerical results of \cite{Albrecht2020} plotted in their figures~6\textit{g}--\textit{i}. The largest component of $\mathbf{F}_\mathrm{NL}$ is the axial component and it exhibits a negative and a positive lobe with a maximum value of $3.6 \times 10^{-3}$, while it is of $5 \times 10^{-3}$ in the simulations. The radial component has a weak dependence on $z$ and the azimuthal component is of order $Ek^{1/2}$ weaker than the other components, except for $\mathbf{F}_\mathit{BL} $ in the boundary layers. 


The azimuthal component of the axisymmetric forcing, $F_\mathrm{az.}=\mathbf{F}\cdot\etheta$, plays a particular role here because it induces a geostrophic flow at an order $\Ek^{-1/2}$ higher than itself. The reason is that $F_\mathrm{az.}$ directly forces a radial flow through the $\theta$-projection of the Navier--Stokes equation, $u_\mathrm{az.} = \frac{1}{2} F_\mathrm{az.}$, which itself forces an axial flow through the incompressibility condition
\begin{equation}
w_\mathrm{az.}(r,z) = -\int \frac{1}{r} \frac{\partial(r u_\mathrm{az.})}{\partial r} \,\rmd z. 
\end{equation}
To cancel the $z$-odd part of this flow on the end walls, a geostrophic flow is necessary such that $(\tilde{w}_0)_\mathrm{az.} = - w_\mathrm{az.}$ in $z=-h/2$. This axial flow then yields a geostrophic flow 
\begin{equation}\label{eq.v0generic}
(v_0)_\mathrm{az.}(r) =\Ek^{-1/2} \int_{-h/2}^0 F_\mathrm{az.} \,\rmd z. 
\end{equation}
The $\theta$-component of the axisymmetric forcing $F_\mathrm{az.}$ therefore induces a geostrophic flow at  order $F_\mathrm{az.}\Ek^{-1/2}$, which is found to be proportional to the $z$-average of $F_\mathrm{az.}$.
Here, we note that this reasoning allows us to show that the non-linear interactions between two Kelvin modes  of different axial wavenumbers do not force a geostrophic flow. For instance, consider a non-resonant Kelvin mode of real amplitude $A_i=O(\Ron)$ and wavenumber $k_i = (2i - 1) \pi/h$ interacting with the resonant Kelvin mode of amplitude $A$ and wavenumber $k = (2j-1) \pi/h$. The non-linear interactions will yield a non-zero azimuthal forcing $F_\mathrm{az.}$ with terms proportional to $\Ron\, \Imag(A) \cos (k\pm k_i) z$. However, using \eqref{eq.v0generic}, we see that this forcing does not give rise to any geostrophic flow because the sum or difference of axial wavenumbers, $(k\pm k_i)$, will always be an integer multiple of $2\pi/h$ and will have a zero average along $z$.

\begin{figure} 
\centerline{\includegraphics[scale=0.75]{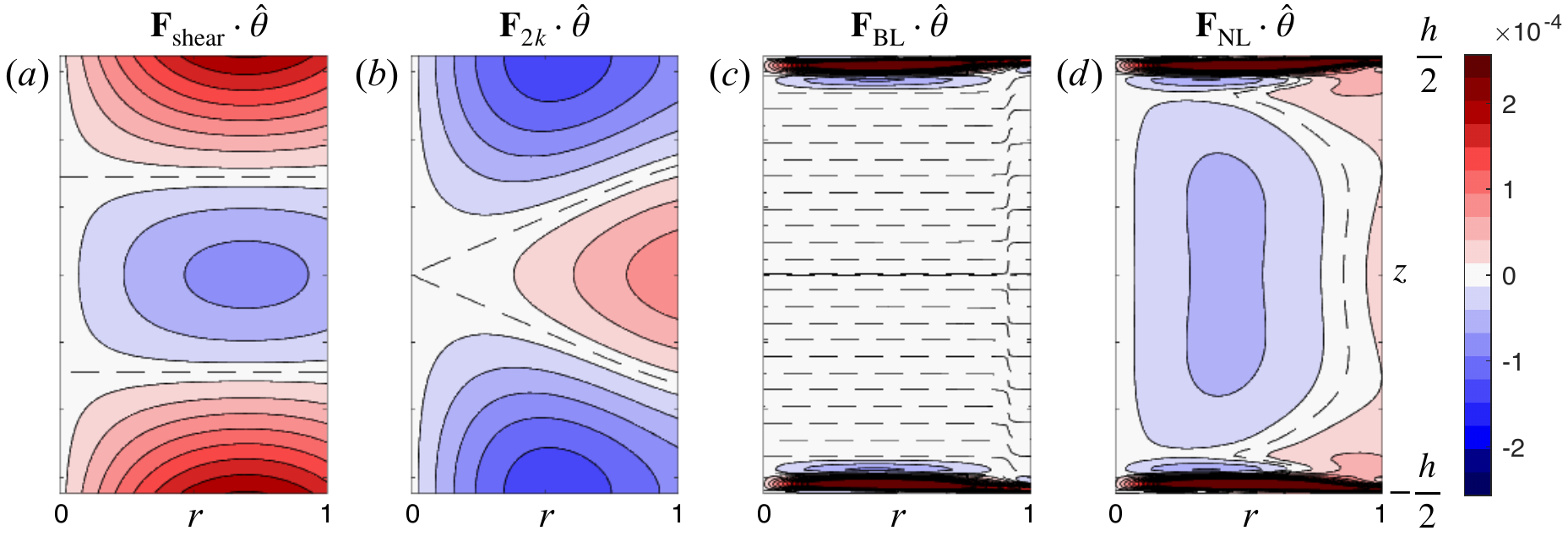}}
\caption{Azimuthal forcing terms of the different components of the geostrophic flow: 
(a) $\mathbf{F}_\mathrm{shear} \cdot \etheta$;
(b) $\mathbf{F}_{2k}           \cdot \etheta$;
(c) $\mathbf{F}_\mathit{BL}     \cdot \etheta$;
(d) $\mathbf{F}_\mathrm{NL}    \cdot \etheta$, with $\mathbf{F}_\mathrm{NL} = \mathbf{F}_\mathrm{shear} + \mathbf{F}_{2k} + \mathbf{F}_\mathit{BL} $.
Parameters are the same as in figure~\ref{f.ForcingRes1}.}
\label{f.ForcingRes1theta}
\end{figure}

The azimuthal components of the different forcing terms are plotted in figures~\ref{f.ForcingRes1}\textit{b},\textit{e} and \ref{f.ForcingRes1theta}\textit{a}--\textit{d}. It shows that $\mathbf{F}_k\cdot\etheta$ and $\mathbf{F}_\mathrm{NL}\cdot\etheta$ are both negative (except near $r=1$ for $\mathbf{F}_\mathrm{NL}\cdot\etheta$), even if the components $\mathbf{F}_\mathrm{shear}\cdot\etheta$ and $\mathbf{F}_\mathit{BL} \cdot\etheta$ have positive average along $z$. 

	

\subsection{Zonal flow}

\begin{figure} 
\centerline{\includegraphics[scale=0.75]{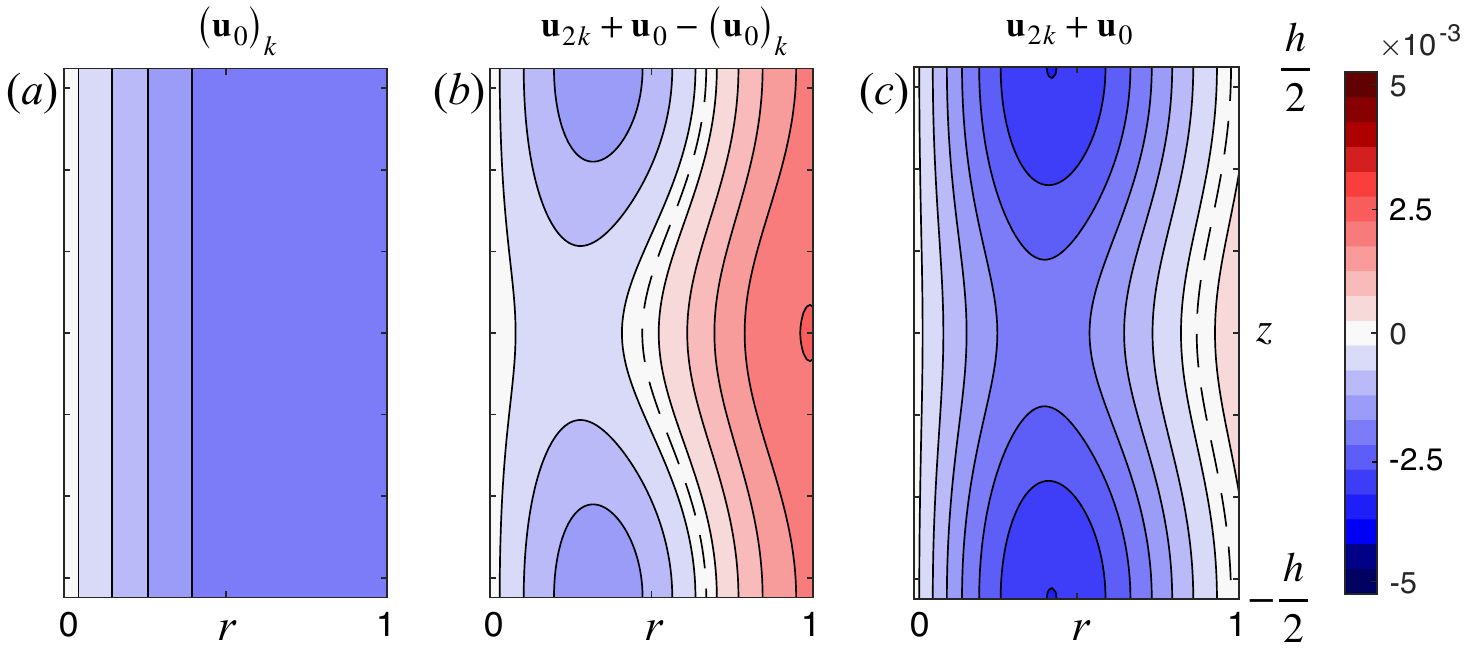}}
\caption{Mean azimuthal flow generated by different sources. 
(a) Velocity $\left(\vit_0\right)_{k}$ given by \eqref{eq.v0k} and induced by the forcing $\mathbf{F}_k$. 
(b) Velocity $\left(\vit_0\right)_\mathrm{shear} + \vit_{2k} + \left(\vit_0\right)_{2k} + \left(\vit_0\right)_\mathit{Ek} + \left(\vit_0\right)_\mathit{BL} = \vit_{2k} + \vit_0 - \left(\vit_0\right)_{k}$ given by \eqref{eq.v0shear}, \eqref{eq.v2k+v02k}, \eqref{eq.w2kbis} and \eqref{eq.v0NL} and induced by the nonlinear forcing term $\mathbf{F}_\mathrm{NL}$. 
(c) Total azimuthal velocity $\vit_{2k} + \vit_0$ induced by all sources. 
Parameters are the same as in figures~\ref{f.ForcingRes1} and \ref{f.ForcingRes1theta}.}
\label{f.ResponseFlowRes1}
\end{figure}

The geostrophic flow $\left(\vit_0\right)_{k}$ produced in response to $\mathbf{F}_k$ is plotted in figure~\ref{f.ResponseFlowRes1}\textit{a}. 
Its azimuthal velocity is always negative with a minimum value equal to $-2.2 \times 10^{-3}$ in good agreement with the minimum value of about $-2.5 \times 10^{-3}$ found numerically by \cite{Albrecht2020} (see their figure 6\textit{o}). However, in the numerical simulations, the azimuthal velocity vanishes at $r=1$ in order to satisfy the no-slip boundary conditions whereas it remains negative in the present theory. 
	
The azimuthal flow $\left(\vit_0\right)_\mathrm{shear} + \vit_{2k} + \left(\vit_0\right)_{2k} + \left(\vit_0\right)_\mathit{Ek} + \left(\vit_0\right)_\mathit{BL} = \vit_{2k} + \vit_0 - \left(\vit_0\right)_{k}$  in response to the nonlinear forcing term $\mathbf{F}_\mathrm{NL}$
is plotted in figure~\ref{f.ResponseFlowRes1}\textit{b}.
It exhibits two negative lobes of azimuthal velocity with a minimal value equal to $-1.6 \times 10^{-3}$.  This value can be compared to the value of $-2.5\times 10^{-3}$ found numerically by \cite{Albrecht2020} (their figure 6s), and the difference can be explained by the 30\% difference in the Kelvin mode kinetic energy.

Adding all responses yields the zonal flow $ \vit_{2k} +
\vit_0$ (figure~\ref{f.ResponseFlowRes1}\textit{c}). It is almost everywhere negative except near the lateral wall around the equator. The velocity distribution is quantitatively similar to that found numerically by \cite{Albrecht2020} in their figure~6k. We find a minimum value of $-3.3 \times 10^{-3}$, close to the numerical value of $-4.5 \times 10^{-3}$ with again a difference that can mainly be imputed to the 30\% difference in kinetic energy.

\begin{figure} 
\centerline{\includegraphics[scale=0.75]{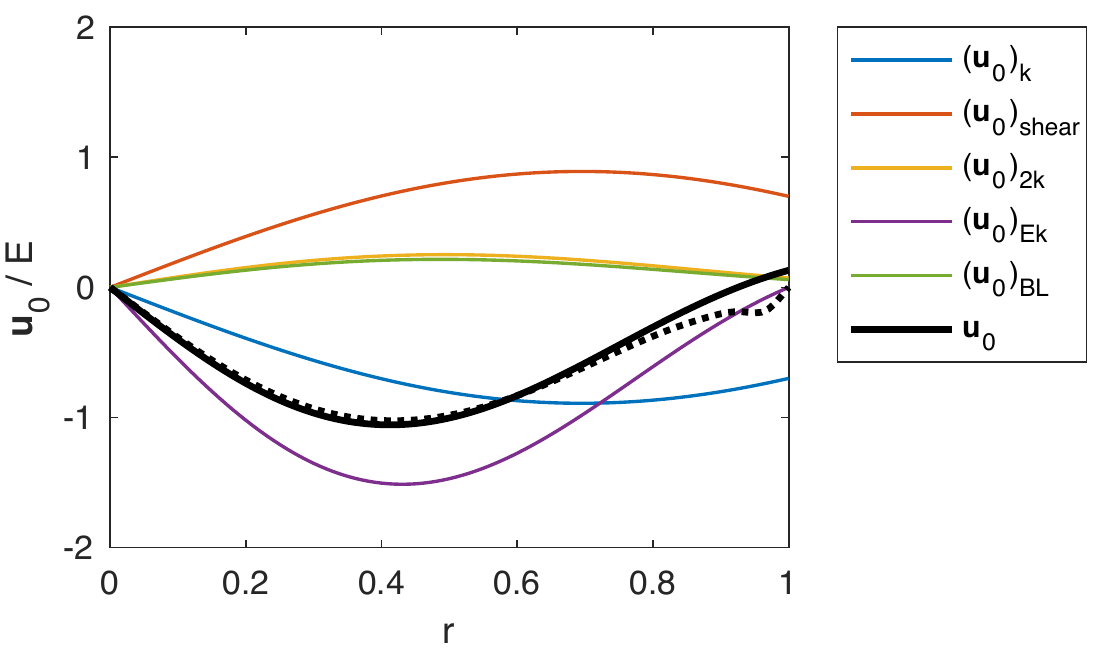}}
\caption{Geostrophic flow generated by different sources. 
The components $\left(\vit_0\right)_{k}$, $\left(\vit_0\right)_\mathrm{shear}$, $\left(\vit_0\right)_{2k}$, $\left(\vit_0\right)_\mathit{Ek}$  and $\left(\vit_0\right)_\mathit{BL} $, given by \eqref{eq.v0k}, \eqref{eq.v0shear}, \eqref{eq.v02k}, \eqref{eq.v0Ek} and \eqref{eq.v0NL}, are represented by solid lines. 
They sum up to form the total geostrophic flow $\vit_0$ (black solid line). The velocity is normalized by the kinetic energy of the Kelvin mode $E$. 
The dotted black line show the numerical results of \cite{Albrecht2020} for the same parameters (see their figure~10).
Parameters are the same as in figures~\ref{f.ForcingRes1} and \ref{f.ResponseFlowRes1}. }
\label{f.vgeo_vs_r}
\end{figure}

Averaging this total flow over the height of the cylinder filters out the non-geostrophic response $ \vit_{2k} $ leading to the total geostrophic flow $ \vit_0$. This geostrophic flow, plotted in figure~\ref{f.vgeo_vs_r} as a black solid line, is in excellent agreement with the numerical result of \cite{Albrecht2020} plotted as a black dotted line. This geostrophic flow contains several terms, the Coriolis term $\left(\vit_0\right)_{k}$ being exactly opposite to the shear term  $\left(\vit_0\right)_\mathrm{shear}$. Apart from these two terms, the dominant contribution comes from the nonlinear coupling between the inviscid Kelvin mode and its viscous correction in the bulk  $\left(\vit_0\right)_\mathit{Ek}$, which is retrograde. This flow is weakly compensated by the geostrophic adaptation $\left(\vit_0\right)_{2k}$ to the zonal flow $vit_{2k}$ and by the nonlinear coupling in the boundary layers $\left(\vit_0\right)_\mathit{BL} $. 
We note here that $\left(\vit_0\right)_{2k}$ and $\left(\vit_0\right)_\mathit{BL} $ are approximately equal for the parameters chosen. This is merely incidental. 

\subsection{Dependence of the zonal flow on the aspect ratio $h$}

We have shown that, in the steady case, the geostrophic flow is proportional to $|A|^2$. This was also found numerically by \cite{Albrecht2020}: in their figure 11c, they showed that the ratio between the kinetic energy of the azimuthal flow $E_0$ and the square of the kinetic energy of the Kelvin mode $E^2$ is nearly independent of the Ekman number and the tilt angle. Here, we show that $E_0$ is proportional to $|A|^4$, while $E$ given by \eqref{eq.E} is proportional to $|A|^2$, which is consistent with the observation of \cite{Albrecht2020}. 

\begin{figure} 
\centerline{\includegraphics[scale=0.75]{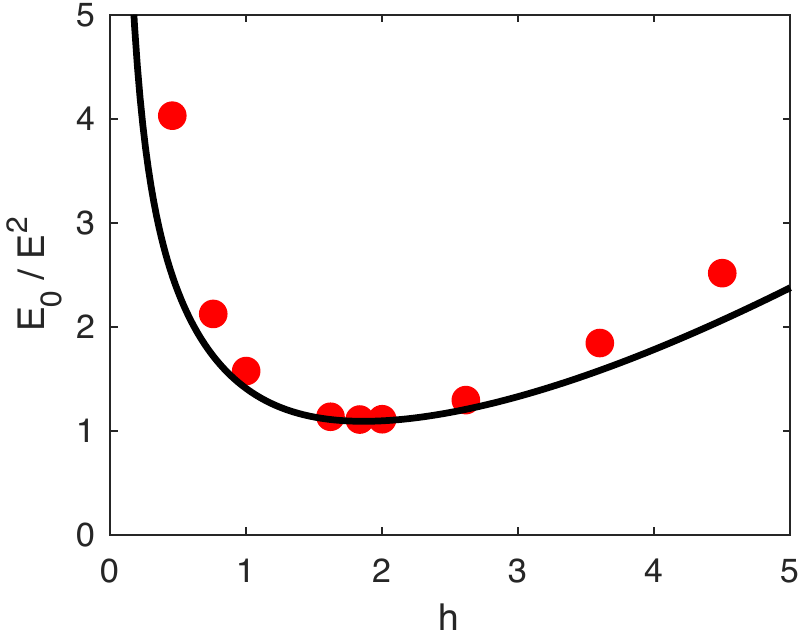}}
\caption{Ratio between the kinetic energy  $E_0$ of the zonal flow $\vit_{2k}+\vit_0$ and the squared kinetic energy $E$ of the forced Kelvin mode. This ratio is plotted as a function of the aspect ratio $h$ for the $(1,1)$ resonance. The symbols correspond to the numerical results obtained by \cite{Albrecht2020} for $\Ek=2\times 10^{-4}$. The solid line corresponds to the present theory.}
\label{f.Amplitude_vs_h}
\end{figure}

Although the ratio $E_0 / E^2$ is independent on the forcing amplitude $\Ron$ and Ekman number $\Ek$, it depends on the aspect ratio $h$. This is because the aspect ratio influences the resonant frequency $\omega_{n,j}$ and the radial wavenumber $\delta_{n,j}$. 
In figure~\ref{f.Amplitude_vs_h}, we compare the quantity $E_0 / E^2$ extracted from \cite{Albrecht2020} to our prediction, when $h$ is varied.
	
The theoretical prediction is found to be in excellent agreement with these numerical results although the theory tends to slightly underestimate the axisymmetric flow in the limit of small and large aspect ratios. This may be due to the presence of the boundary layers on the cylinder side walls.

\section{Comparison with experimental results} \label{Sec.ComparisonExperiments}

We shall now compare the present theoretical predictions with the experimental results of \cite{Meunier2008}.

\begin{figure} 
\centerline{\includegraphics[scale=0.75]{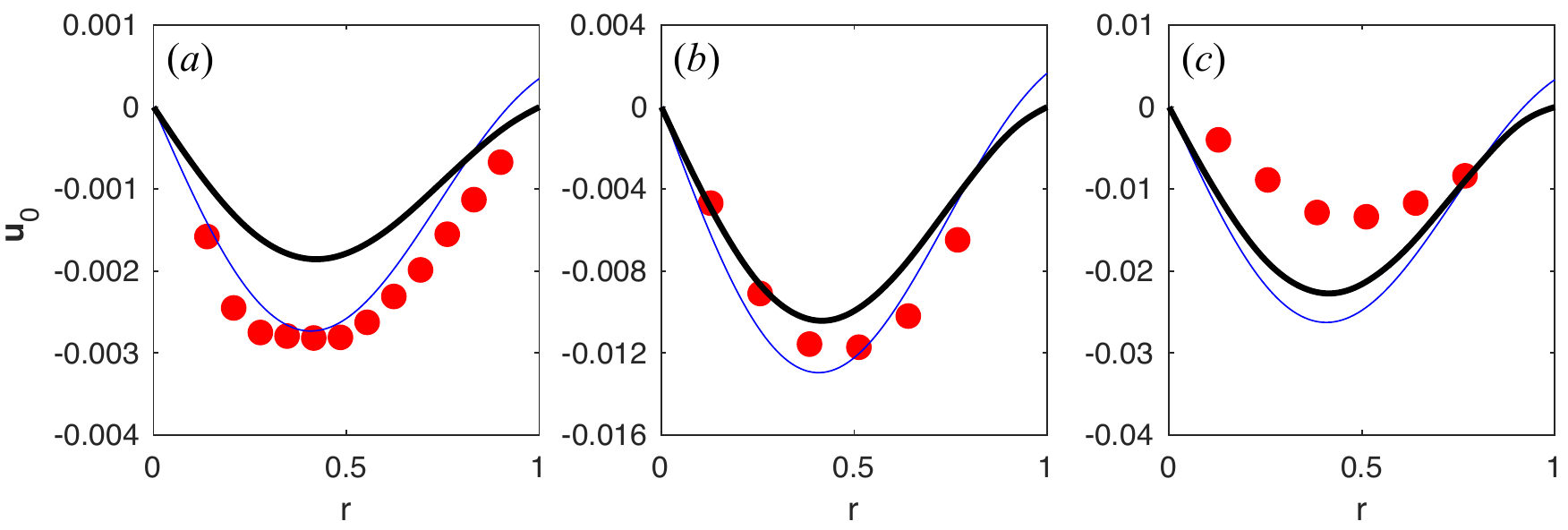}}
\caption{Mean azimuthal velocity  profiles of the $(1,1)$ resonance.
Symbols are the experimental results of \cite{Meunier2008} measured at $z=h/4$, while the solid lines represent the velocity profiles predicted: $|A|^2 v_0$ given by \eqref{eq.v0totr} in blue and $|A|^2 \left(v_0\right)_\mathrm{vol.}$ given by \eqref{eq.v0totrvol} in black.
Parameters are: $h=1.8$, $\omega=1.088$, $ \Ron=-0.0031$ and three different Ekman numbers: (a) $\Ek =8.1 \times 10^{-4}$; (b) $\Ek = 2.0 \times 10^{-4}$; and (c) $\Ek =0.81 \times 10^{-4}$. 
}
\label{f.ExpRes1}
\end{figure}

We start with the $(1,1)$ resonance. The aspect ratio is $h=1.8$, giving a resonant frequency $\omega_{1,1}=1.088$ (table~\ref{tab_NL}).
The tilt angle is $\varphi=2^\circ$, which corresponds to a forcing amplitude $\Ron=-0.0031$. 
The mean azimuthal velocity has been measured for three moderate Ekman numbers at $z=h/4$ (figure~\ref{f.ExpRes1}). In each case, the azimuthal velocity is negative meaning that the zonal flow is retrograde. 	
	
We first compute the amplitudes of the Kelvin mode and geostrophic flow by looking for the fixed point $(A,A_0)$ of the weakly nonlinear amplitude equations \eqref{eq:weaklyNLamplitudeEq}. 
We then plot the resulting geostrophic flow $\vit_0$ for two cases:  
first we consider the profile $v_0$ obtained in section \ref{Sec.ForcingGeostrophic}  and given by \eqref{eq.v0totr}, second we take into account viscous effects onto the geostrophic flow by proceeding as follows.
We decompose $v_0(r)$ into Bessel functions of the first kind
\begin{equation}\label{eq.v0_J1}
v_0(r) = \sum_{i=1}^{\infty} A_i J_1(d_i r), \quad
\mbox{such that }
J_1(d_i) = 0.
\end{equation}
where the amplitudes $A_i$ can be calculated as
\begin{equation}
A_i = \frac{2}{J_0^2(d_i)} \int_{0}^{1} v_0(r) J_1(d_i r)\, r \,\rmd r.
\end{equation}
This decomposition is analogue to a Fourier decomposition. 
The $d_i$'s  are the zeros of $J_1$ ordered in ascending order. They play the role of radial wavenumbers for the geostrophic flow and satisfy $d_i \sim i \pi$ for large $i$. 

When the geostrophic flow $v_0$ is decomposed as in \eqref{eq.v0_J1}, the 
 viscous term of the Navier--Stokes equations, $\Ek \,\nabla^2 \vit_0$, gives rise to an azimuthal forcing $F_\mathrm{az.} = - \sum_i A_i d_i^2 J_1(d_i r)$, which itself yields a geostrophic flow $ \Ek^{1/2} F_\mathrm{az.} h/2$ through the reasoning explained above in \S\ref{subsec.compnum}. 
The corrected azimuthal flow taking into account this  effect is then given by 
\begin{equation}\label{eq.v0totrvol}
\left(v_0\right)_\mathrm{vol.}(r) = \sum_{i=1}^{\infty} \frac{A_i}{1+\Ek^{1/2} d_i^2 h/2} J_1( d_i r).
\end{equation}

This way to incorporate  viscous  effects  has been first introduced in \cite{Meunier2008}. 
It accounts for both the side wall boundary layer and the Ekman pumping from end walls.  
In the limit of small $\Ek$, we recover at leading order the $O(\Ek^{1/4})$ side wall boundary layer obtained by \cite{Wang1970} (see appendix~\ref{appB}). The $O(Ek^{1/2})$ correction 
induced by the Ekman pumping in the bulk depends however on the Bessel decomposition \eqref{eq.v0_J1} of $v_0$. This correction is the smallest if the spectral content of $v_0$ is mainly on small 
radial wavenumbers. For the geostrophic flows plotted in figure~\ref{f.ExpRes1}, the smallest wavenumber $d_1$ is dominant. Viscous effects are therefore expected to be negligible as
soon as the Ekman number is asymptotically small compared to  $d_1^{-4} h^{-2}\approx 10^{-3}$.

This is the case in figure~\ref{f.ExpRes1}c and in the numerical simulations shown in the previous section, but not for the case of figures~\ref{f.ExpRes1}a. Note that, in the limit $\Ek \ll d_1^{-4} h^{-2}$, the flow is generally  unstable unless the tilt angle is extremely small, which is difficult to achieve experimentally.
		
Comparing the experimental velocity profiles and the theoretical predictions in figure~\ref{f.ExpRes1} shows a good agreement. Both profiles have the same bell shape with a minimum value reached around $r\approx 0.4$. The amplitude is underpredicted by 30\% for the lower Ekman number and overpredicted by a factor almost 2 for the highest value. This latter discrepancy may be due to large nonlinear effects and an instability of the Kelvin mode.

\begin{figure} 
\centerline{\includegraphics[scale=0.75]{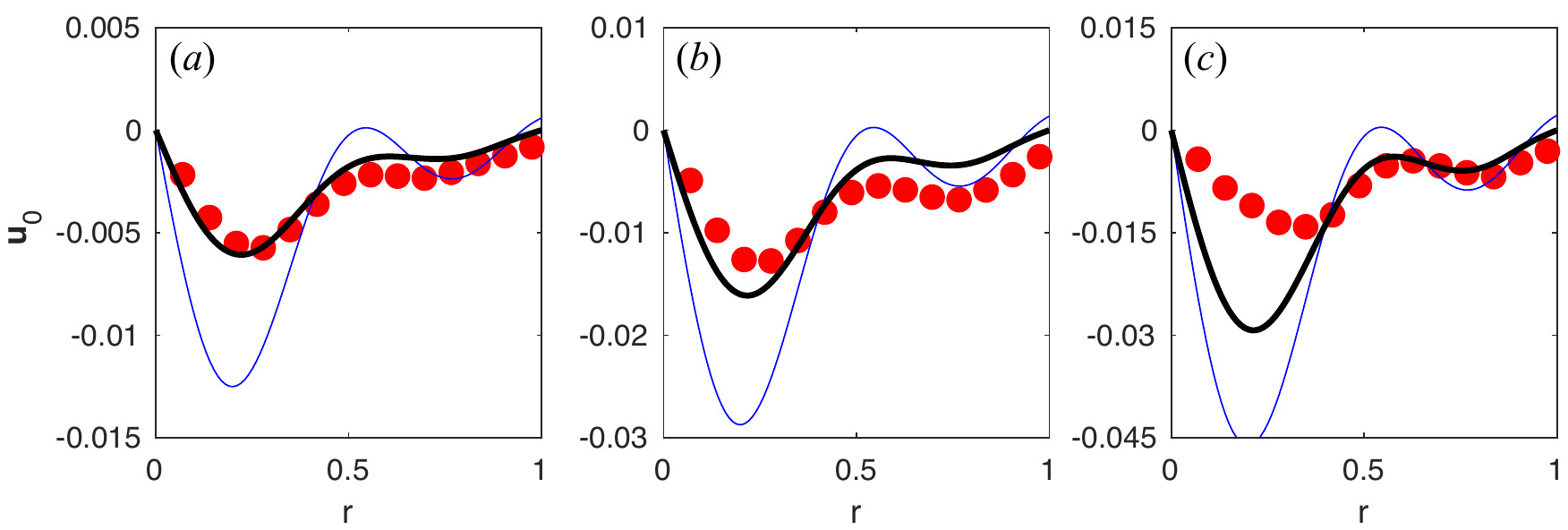}}
\caption{Same as figure~\ref{f.ExpRes1} for the $(2,1)$ resonance.
Parameters are: $h=2$, $\omega=0.510$, $\Ron=-0.0086$, $z=h/4$ and (a) $\Ek =2.66 \times 10^{-4}$; (b) $\Ek = 1.33 \times 10^{-4}$; and (c) $\Ek =0.66 \times 10^{-4}$. 
}
\label{f.ExpRes2}
\end{figure}

The comparison for the $(2,1)$ resonance is shown in figure~\ref{f.ExpRes2}. In this case, the viscous effects on the geostrophic flow are even more pronounced than for the $(1,1)$ resonance in figure~\ref{f.ExpRes1}. This is because of the importance of the second  amplitude $A_2$ in the Bessel series  \eqref{eq.v0_J1}. The agreement between theory and experiment is excellent both qualitatively and quantitatively. Both  profiles are negative with two local minima: the largest one is around $r=0.25$ and the second one around $r=0.75$. The amplitudes predicted by $\left(v_0\right)_\mathrm{vol.}$ compare well with experimental results for the two largest values of the Ekman number. For the smallest Ekman number, as for the first resonance, the discrepancy may be due to nonlinearities or an instability. 
	
\section{Retrograde sign of the geostrophic flow}
\label{Sec.RetrogradeDirection}

So far, we have found that the geostrophic flow is always retrograde. This is a classical observation in precessing flows, where precession tends to slow down the solid body rotation \citep{Kobine1996,Meunier2008}. Although this  seems intuitive, there is no proof of this result so far. In fact, nothing prevents precession from spinning up the solid body rotation because the precessional motion injects energy in the system. For instance, this is what would happen if only the nonlinear term $(\vit_0)_\mathit{BL} $ was considered for the $(1,1)$ resonance (figure~\ref{f.vgeo_vs_r}). 

\begin{figure} 
\centerline{\includegraphics[scale=0.75]{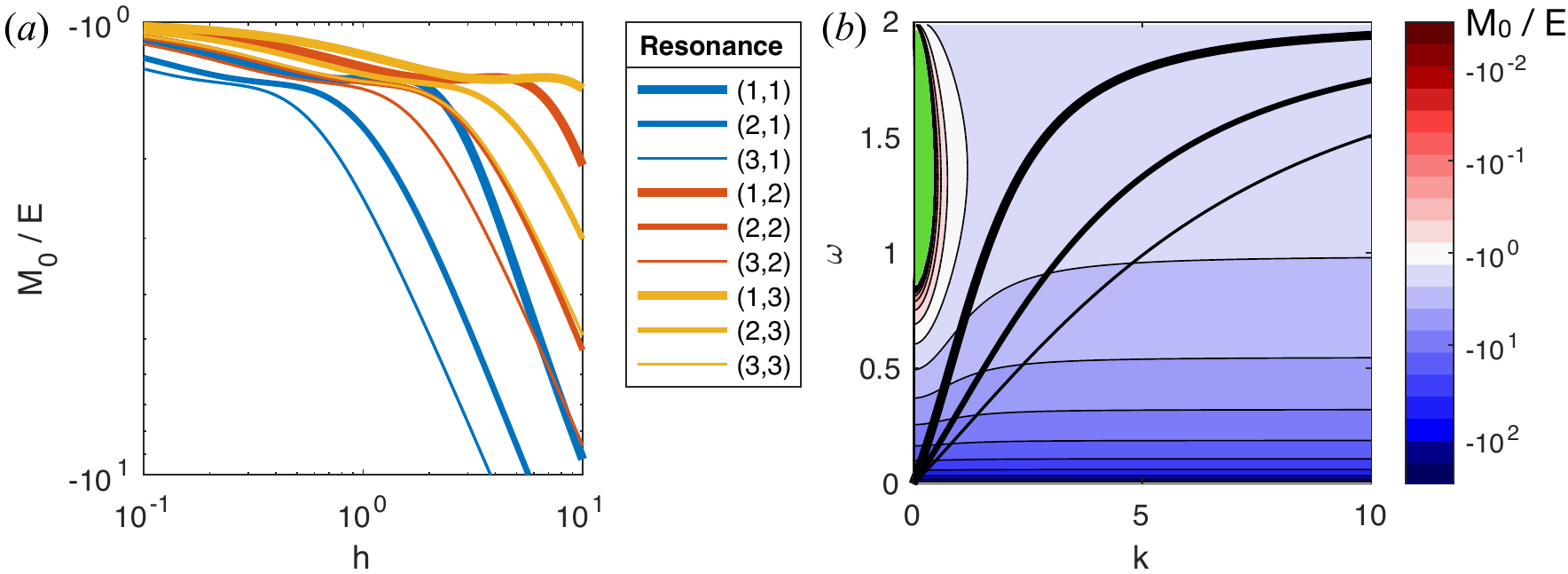}}
\caption{
Angular momentum $M_0$ of the geostrophic flow normalised by the kinetic energy  of the Kelvin mode $E$ for a resonant precessing cylinder ($m=1$). 
In (a), the angular momentum is plotted as a function of the aspect ratio $h$ for different resonances as mentioned in the legend.  
In (b), the angular momentum is plotted as contours in the $(k,\omega)$-plane. The green zone corresponds to positive values of the angular momentum, but this zone is not accessible due to the dispersion relation. 
The black lines show this dispersion relation for the first three resonances: $(1,\cdot)$ thick line; $(2,\cdot)$ medium line; and $(3,\cdot)$ thin line (the higher resonances have lower $\omega$). }
\label{f.AngularMomentum}
\end{figure}

To assess whether the zonal flow $\vit_0 + \vit_{2k}$ is always retrograde, we calculate its angular momentum defined as
\begin{equation}\label{eq.M0}
M_0 = \ez \cdot \int_V \pos \times \left(\vit_0 + \vit_{2k}\right)\, \rmd V 
	= 2 \pi h A_0 \int_{0}^{1} r^2 v_0(r)\, \rmd r ,
\end{equation}
which in fact only depends on the geostrophic component $\vit_0$, the non-geostrophic part $\vit_{2k}$ being periodic along $z$. 
In figure~\ref{f.AngularMomentum}\textit{a}, we see that this angular momentum (normalized by the kinetic energy of the Kelvin mode $E$) is always negative for the range of aspect ratios and resonances chosen. 	

The angular momentum $(M_0)_\mathit{BL}$ due to the component $(v_0)_\mathit{BL}$ of the geostrophic flow is complex, but if we first focus on the components $(v_0)_{2k}$ and $(v_0)_\mathit{Ek}$ of the geostrophic flow, we can show that their associated angular momenta simplify into 
\begin{eqnarray}    
\frac{\left(M_0\right)_{2k}}{E} & = & 1 - m\frac{
   \omega k^2 + m (m \omega  - 2)} 
   {2 k^2 + m (2m - \omega)} .\\
\frac{\left(M_0\right)_\mathit{Ek}}{E} & = & - \frac{h m} {k} \,
	\Imag \left(k^{(1)}\right) 
	= - \frac{m}{\sqrt{2}\,\omega} \left(
		\frac{2+\omega}{(2 - \omega)^{1/2}} +
		\frac{2-\omega}{(2 + \omega)^{1/2}}\right) 	
	,    
\end{eqnarray}
where $E$ is the kinetic energy of the Kelvin mode given by \eqref{eq.E}.
The ratio $(M_0)_\mathit{Ek} / (mE)$ is a function of $\omega$ only that is always negative on the interval $0 < \omega < 2$. With some calculation, it can also be shown that the sum $(M_0)_{2k} + (M_0)_\mathit{Ek}$ is always negative for $m \ge 1$, $k>0$ and $0 < \omega < 2$.  

To show numerically that the total angular momentum is always retrograde, we use the fact that the ratio $(M_0)_\mathit{BL} / E$ is a function of $k$ and $\omega$ only.
Although this function is too complex to reproduce here, it can be easily plotted. 
We can thus plot the total angular momentum $M_0 / E$ in the $(k,\omega)$-plane for $m=1$ (figure~\ref{f.AngularMomentum}\textit{b}).
In this plane, we see that the angular momentum is everywhere negative, except in a small region for small $k$ and $0.8\lesssim\omega<2$ (green region in figure~\ref{f.AngularMomentum}\textit{b}).
However, this region is not reachable by any resonance, because $k$ and $\omega$ are linked by the dispersion relation  $u(1)=0$ with $u$ given in (\ref{uvw}\textit{a}). This dispersion relation is plotted as black lines in figure~\ref{f.AngularMomentum}\textit{b} for the first three resonances. For higher resonances, the angular frequency continue to decrease and moves further aways from the zone of positive angular momentum. It thus shows that the angular momentum is always retrograde for $m=1$ and $0<\omega<2$.

\begin{figure} 
\centerline{\includegraphics[scale=0.75]{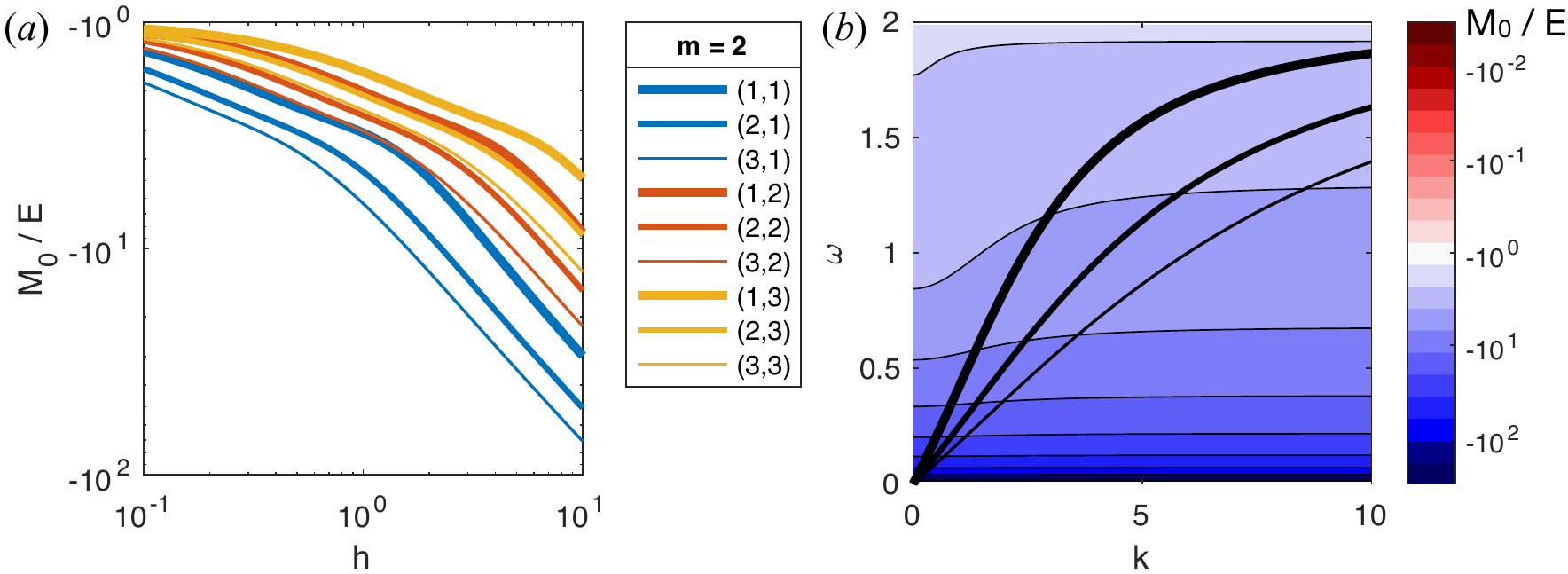}}
\caption{Same as figure~\ref{f.AngularMomentum} for the Kelvin modes of azimuthal wavenumber $m=2$. }
\label{f.AngularMomentum_m2}
\end{figure}

Although the calculations presented in this paper focused on Kelvin modes of azimuthal wavenumber $m=1$ forced by precession, they are easily generalisable to different azimuthal wavenumbers. 
In figure~\ref{f.AngularMomentum_m2}, we consider the azimuthal wavenumber $m=2$.
It shows again that a resonant Kelvin modes always forces a retrograde geostrophic flow. 
In this case, it is even simpler because, contrarily to $m=1$, there is no region of positive angular momentum in the $(k,\omega)$-plane.

%
\begin{figure} 
\centerline{\includegraphics[scale=0.75]{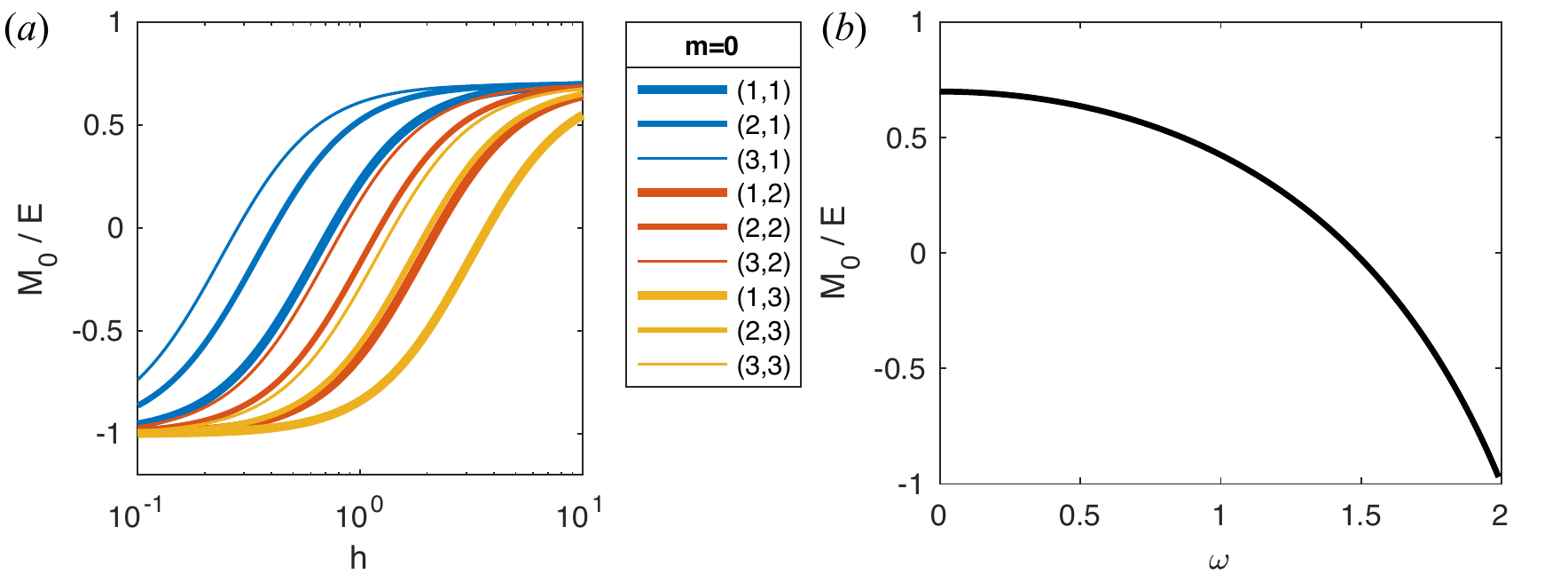}}
\caption{Same as figure~\ref{f.AngularMomentum} for the Kelvin modes of azimuthal wavenumber $m=0$. In this case, the ratio between the angular momentum $M_0$ and the kinetic energy $E$ is only a function of $\omega$ represented in (b). It changes sign for $\omega \approx 1.48$.}  
\label{f.AngularMomentum_m0}
\end{figure}

The situation is different for the azimuthal wavenumber $m=0$ (figure~\ref{f.AngularMomentum_m0}). In this case, the sign of the angular momentum $M_0$ is positive for $\omega \lesssim 1.48$ and negative for $\omega \gtrsim 1.48$ (figure~\ref{f.AngularMomentum_m0}\textit{b}). It thus means that the geostrophic flow will be retrograde in the limit of small aspect ratio but prograde in the limit of large aspect ratio (figure~\ref{f.AngularMomentum_m0}\textit{a}).

The figures~\ref{f.AngularMomentum}\textit{b} and \ref{f.AngularMomentum_m2}\textit{b} show numerically that the geostrophic flow is always retrograde for azimuthal wavenumbers $m=1$, $2$ and for forcing frequency in the interval $0<\omega<2$. 
The situation is similar for $m=3$ (not shown here) and presumably for higher values of the azimuthal wavenumber.
But again, this is difficult to prove mathematically given the complexity of  $(M_0)_\mathit{BL}$. 

Note that for negative forcing frequency ($\omega <0$), the geostrophic flow could be prograde. 
This corresponds to the particular case of a cylinder rotating around its axis with a frequency $0 < \Omega_0 < -\Omega_p$, with $\Omega_p < 0$ the precession frequency. 
Usually, experiments and numerical simulations do not consider this limit and rather focus on proper frequencies larger than precession frequencies ($|\Omega_0| > |\Omega_p|$).

\section{Conclusion}
\label{Sec.Conclusion}

In this paper, we have provided an analytic expression of the zonal flow in a precessional cylinder at resonance. In this regime, the harmonic response is dominated by a single Kelvin mode of amplitude $A$, azimuthal
wavenumber $m=1$ and axial wavenumber $k$. 
We have identified five contributions to the zonal flow.

The first contribution, noted $ \left(\vit_0\right)_{k}$, and given by \eqref{eq.v0k}, comes from the interaction of the Kelvin mode with the equatorial Coriolis force.  
 The second contribution comes from the interaction of the Kelvin mode with the axial shear (i.e. the particular solution to the Poincar\'e forcing). Surprisingly, its expression is exactly opposite to 
  $ \left(\vit_0\right)_{k}$ whatever the Kelvin mode.  As a consequence, these two first contributions, which are specific to the precessing cylinder, cancel each other. 
  
 The three other contributions come from the interaction of the Kelvin mode with itself. They are therefore not specific to the precessing cylinder and their expressions have been 
 provided for any $k$ and $m$. 
The first one corresponds to the steady and axisymmetric term $\vit_{2k}$ of amplitude $|A|^2$ and axial wavenumber $2k$, which is forced by the nonlinear interaction of the inviscid Kelvin mode with itself. 
This inviscid (non-geostrophic) zonal flow exhibits a non-zero azimuthal velocity on the end walls. It must be complemented by a geostrophic flow $ \left(\vit_0\right)_{2k}$ to satisfy the no-slip boundary condition. The sum of these two terms is given analytically in \eqref{eq.v2k+v02k}.

The second Kelvin mode source of zonal flow comes from the nonlinear interaction of the inviscid Kelvin mode with its viscous correction in the bulk. This weak forcing of order $|A|^2 \Ek^{1/2}$ generates an axial flow of order $|A|^2 \Ek^{1/2}$ with non-zero velocity at the end walls. Such a flow must thus be compensated by the Ekman pumping of a geostrophic flow, $\left(\vit_0\right)_\mathit{Ek}$, of order $|A|^2$ given in \eqref{eq.v0Ek}. 

 Finally, the third Kelvin mode contribution to the zonal flow comes from the nonlinear self-interaction of the flow inside the end wall boundary layers. It generates a non-zero axial flow at the end walls, of order $|A|^2 \Ek^{1/2}$, which again must be compensated by a geostrophic flow of order $|A|^2$, given analytically in \eqref{eq.v0NL}. 
 
We have used these expressions to derive the coupled equations that describe the slow dynamics of the Kelvin mode amplitude and the geostrophic flow amplitude. 
These equations also provide the saturation amplitude of the Kelvin mode. 
The variation of the nonlinear coefficient coming from the interaction with the zonal flow has been analysed as a function of the cylinder aspect ratio for the first nine resonances. 

The present results have been compared to the numerical simulations of \cite{Albrecht2020} and to the experimental measurements of \cite{Meunier2008}. A good agreement has been observed, especially when
the viscous correction to the zonal  flow was considered.  

We have also computed the angular momentum of the zonal flow to assess whether it is prograde (positive angular momentum) or retrograde (negative angular momentum).  
This  has allowed us to show numerically that the zonal flow is always retrograde for $m=1$, $2$, or $3$, and presumably for larger $m$. 

The zonal flow has been calculated for any Kelvin mode. 
Our results can thus be applied to other azimuthal forcing, such as libration.
This  $m=0$ forcing has been studied in a cylinder by \cite{Wang1970,Noir2010,Busse2010,Lopez2011,Sauret2012}.  \cite{Wang1970} provided an expression for the zonal flow far from resonant conditions.
The present work is expected to apply when a Kelvin mode is resonantly excited.  If this mode is the dominant harmonic response, the three  contributions that have been calculated here can be used 
to compute the zonal flow. 
Contrarily to precession,  we have seen that the  zonal flow generated by Kelvin modes with $m=0$ is not necessarily retrograde. We have in particular shown that it becomes prograde for large aspect ratios.

Our results should also be useful to describe the dynamics of interacting Kelvin modes.
Having a good description of the zonal flow they generate is indeed essential to predict correctly the evolution of their amplitudes. 
It would therefore be interesting to revisit, in light of the present results, the weakly nonlinear analyses that have been published on the 
elliptic instability \citep{Waleffe,Mason1999,Eloy2003} and on other parametric instabilities in a cylinder \citep{kerswell1999,Racz2008b,Lagrange2011,lagrange2016triadic}. 
Our calculation gives a way to estimate the geostrophic feedback. This would certainly help to gain insight into  the transition  between wave turbulence and geostrophic turbulence 
when  the Ekman number is increased \citep{LeReun2019}. 

Our analysis has considered Kelvin modes in a cylinder but a similar approach can be developed in a sphere. 
Kelvin modes are  well known in this geometry \citep{Bryan1889,Greenspan1968} and they can also  be resonantly excited \citep{Aldridge1969}.
The zonal flow that is generated by their interaction in the bulk and in the boundary layer can be calculated by the same method. It gives us a hope
to possibly predict the complex zonal flow that has been observed in this geometry  when  a Kelvin mode is resonantly excited \citep{Morize2010}.
Note however that for other geometries as the spherical shell, one would have to consider the presence of internal shear layers \citep{Kerswell1995,LeDizes2017} and 
attractors \citep{Rieutord1997,Rieutord2018} in the harmonic response to possibly describe the induced zonal flow \citep{Tilgner2007,Favier2014,Lin2020,LeDizes2020}.

\begin{acknowledgments}
We are grateful to Hugh Blackburn and Thomas Albrecht for providing their numerical results. 
Donglai Gao thanks the Chinese Science Council for financing a two-year scholarship. 

Declaration of Interests. The authors report no conflict of interest.
\end{acknowledgments}

\appendix
\section{Coefficients of nonlinear interactions}\label{appA}
The functions needed to express the total velocity flow appearing in \eqref{eq.utotBL} are 
\begin{subequations}\label{eq.uvwtot}
\begin{align}
u_\mathrm{tot.} & = u_1 \rme^{-\kappa_1 \tilde z} + u_2 \rme^{-\kappa_2 \tilde z} + u,\\
v_\mathrm{tot.} & = u_1 \rme^{-\kappa_1 \tilde z} - u_2 \rme^{-\kappa_2 \tilde z} + v,\\
w_\mathrm{tot.} & = w_1 \rme^{-\kappa_1 \tilde z} + w_2 \rme^{-\kappa_2 \tilde z} + w_3 + w_4 \tilde z, \label{eq.uvwtotc}
\end{align}
\end{subequations}
with 
	\begin{equation}\label{eq.k1k2}
		\kappa_1 = \left(1+\rmi\right) \sqrt{1+\omega/2}, \quad
		\kappa_2 = \left(1-\rmi\right) \sqrt{1-\omega/2}.
	\end{equation}
and the radial functions
	\begin{subequations}\label{eq.u12v12w1234}
		\begin{align}
		u_1&=-(u+v)/2,\\
		u_2&=-(u-v)/2,\\
		w_1&=((1+m)u_1 + ru'_1)/(r\kappa_1), \\
		w_2&=((1-m)u_2 + ru'_2/(r\kappa_2),\\
		w_3&=-w_1 - w_2, \\
		w_4&=-k w, 
		\end{align}
	\end{subequations}
where  $u$, $v$ and  $w$ are given in \eqref{uvw}.

The scalars $\kappa_i$ 
needed to compute $\tilde{\qvit}_\mathit{BL} $ in \eqref{eq.v0NL} are given by \eqref{eq.k1k2} and
	\begin{equation}
	\kappa_3=\bar{\kappa}_1+\kappa_1,\quad 
	\kappa_4=\bar{\kappa}_2+\kappa_2,\quad 
	\kappa_5=\bar{\kappa}_1+\kappa_2,
	\end{equation}
The radial functions $a_i$ and $b_i$ are given by 
	\begin{subequations}
		\begin{align}
		a_1&=u_1(\kappa_1 \bar w_3 - u' + (m+2) v/r) + u(mu_1/r - u'_1),\\
		a_2&=u_2(\kappa_2 \bar w_3 - u' + (m-2) v/r) - u(mu_2/r + u'_2),\\
		a_3&=u_1\left(\kappa_1 \bar w_1 + u'_1 -{4 \kappa_1 u'_1}/{\kappa_3} + {(1+m)u_1}/{r}\right),\\
		a_4&=u_2\left(\kappa_2 \bar w_2 + u'_2 -{4 \kappa_2 u'_2}/{\kappa_4} + {(1-m)u_2}/{r}\right),\\
		a_5&=\bar{\kappa}_1 u_1 w_2  + \kappa_2 u_2 \bar w_1 - {2 u_1 u_2}/{r} - u_1 u'_2 - u'_1 u_2,\\
		b_1&=-u_1(\kappa_1 \bar w_3 + v' +  v/r) + u(u_1/r + u'_1),\\
		b_2&= u_2(\kappa_2 \bar w_3 - v' -  v/r) - u(u_2/r + u'_2),\\
		b_3&=u_1\left(-\kappa_1 \bar w_1 + u'_1 + {(m+1)u_1}/{r} \right),\\
		b_4&=u_2\left( \kappa_2 \bar w_2 - u'_2 + {(m-1)u_2}/{r} \right),\\
		b_5&=a_5,
		\end{align}
	\end{subequations}
and the radial functions $c_i$, $d_i$ by
	\begin{subequations}
		\begin{align}
		c_1&=\kappa_1 u_1 w_4;\quad
		c_2=\kappa_2 u_2 w_4;\quad
		c_3=c_4=c_5=0,\\
		d_1&=-\kappa_1 u_1 w_4;\quad
		d_2=\kappa_2 u_2 w_4;\quad
		d_3=d_4=d_5=0,
		\end{align}
	\end{subequations}
where radial dependence is expressed through the functions $u_i$, $v_i$ and $w_i$ given in \eqref{eq.u12v12w1234}. 
	
The forced flow in the lower boundary layer $\tilde\qvit_\mathit{BL} $ is expended in powers of $\Ek^{1/2}$ as 
$\tilde\qvit_\mathit{BL}  = |A|^2 \tilde{\qvit}_\mathit{BL} ^{(0)} + |A|^2\Ek^{1/2} \tilde{\qvit}_\mathit{BL} ^{(1)} + \cdots$. At leading order, the solution is 
	\begin{subequations}\label{eq.vBL0sol}
		\begin{align}
		\tilde{u}_\mathit{BL} ^{(0)} & = 
		\sum_{i=1}^{5} 
		\left(\alpha_i + \check{\alpha}_i \tilde{z}\right) \rme^{-\kappa_i \tilde{z}} -
		\frac{\alpha_i + \rmi \beta_i}{2} \rme^{-(1+\rmi)\tilde{z}} -
		\frac{\alpha_i - \rmi \beta_i}{2} \rme^{-(1-\rmi)\tilde{z}}, \label{eq.vBL0sola}\\
		\tilde{v}_\mathit{BL} ^{(0)}  & = 
		\sum_{i=1}^{5}
		\left(\beta_i + \check{\beta}_i \tilde{z}\right) \rme^{-\kappa_i \tilde{z}} +
		\frac{\rmi \alpha_i - \beta_i}{2} \rme^{-(1+\rmi)\tilde{z}} -
		\frac{\rmi \alpha_i + \beta_i}{2} \rme^{-(1-\rmi)\tilde{z}},\\
		\tilde{w}_\mathit{BL} ^{(0)}  & = 0,
		\end{align}
	\end{subequations}
where the first term in the sum correspond to the particular solution of the forced Navier-Stokes equations and the second and third terms are solutions of the homogeneous equations and ensure that $\tilde{u}_\mathit{BL} ^{(0)}$ and $\tilde{v}_\mathit{BL} ^{(0)}$ are zero in $\tilde z = 0$ and $\infty$. The functions $\alpha_i$, $\check{\alpha}_i$, $\beta_i$ and $\check{\beta}_i$ depend on $r$ and are given by
\begin{subequations}
	\begin{align}
	\alpha_i  &= \frac
	{ a_i \kappa_i^2 (\kappa_i^4+4) - 2 c_i \kappa_i (\kappa_i^4-4) + \rmi \left(2b_i (\kappa_i^4 + 4) - 8 d_i \kappa_i^2\right)}
	{(\kappa_i^4 + 4 )^2},\\
	\check{\alpha}_i &= \frac{-c_i \kappa_i^2+2 \rmi d_i}{\kappa_i^4 + 4 },\\
	\beta_i   &= -\frac
	{2 a_i (\kappa_i^4+4) + 8 c_i \kappa_i^2  +\rmi \left(b_i \kappa_i^2(\kappa_i^4+4) + 2 d_i \kappa_i(\kappa_i^4-4)\right)}
	{(\kappa_i^4 + 4 )^2}
,\\
	\check{\beta}_i  &= -\frac{2 c_i+\rmi d_i \kappa_i^2}{\kappa_i^4 + 4 }.
	\end{align}
\end{subequations}
At next order, the incompressibility condition yields
	\begin{equation}
	\tilde{w}_\mathit{BL} ^{(1)}(r,\tilde{z}) = -\frac{1}{r} \int{ \frac{\partial \left( r \tilde{u}_\mathit{BL} ^{(0)}\right)}{\partial r} \,\rmd \tilde{z}}, 
	\end{equation}
which can be simply derived from \eqref{eq.vBL0sola}.

\section{Viscous effects on the geostrophic flow}
\label{appB}

In this appendix, we compare $\left(v_0\right)_\mathrm{vol.}$,  our viscous approximation for the geostrophic mode given in \eqref{eq.v0totrvol}, with the solution 
\begin{equation}\label{eq.v0Wang}
\left(v_0\right)_\mathrm{Wang}(r) = 
	v_0 (r) \left(1 - \rme^{-(1 - r) / d_\mathit{BL} } \right),
	\quad\mbox{with }
	d_\mathit{BL}  = Ek^{1/4} \sqrt{h/2},
\end{equation}
obtained by \cite{Wang1970} using a boundary layer approach.

\begin{figure} 
\centerline{\includegraphics[scale=0.75]{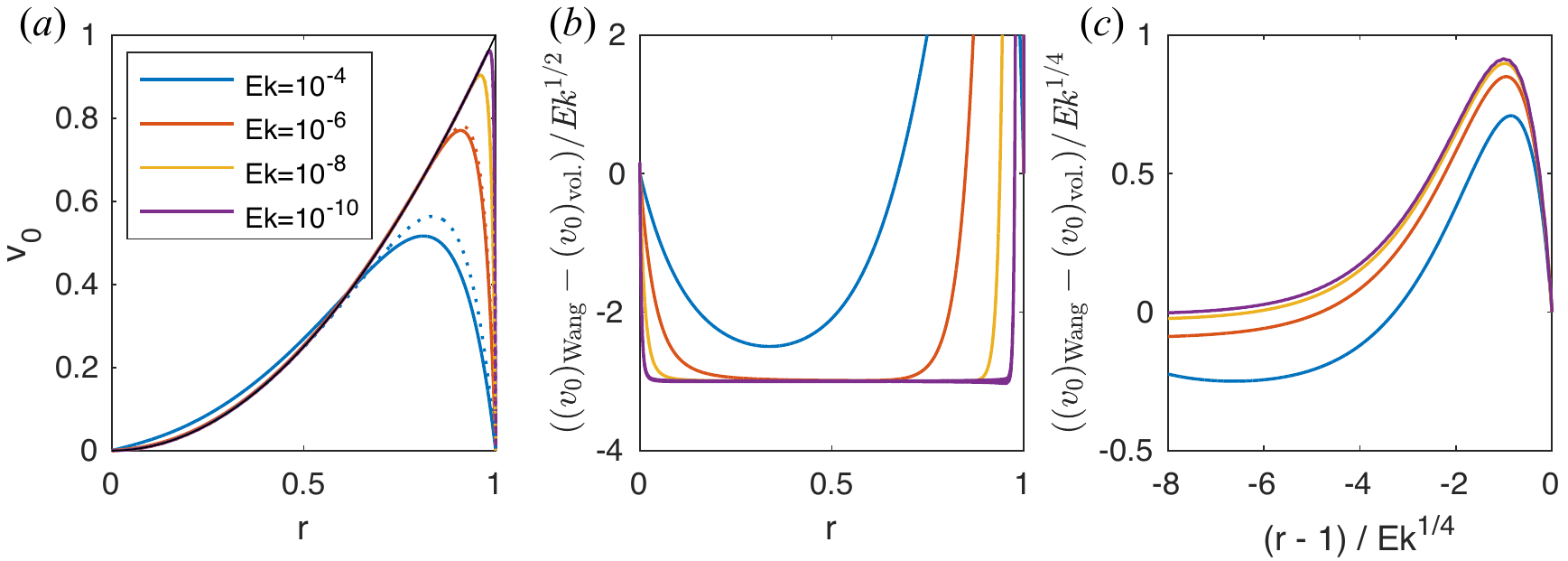}}
\caption{
Viscous effects on the geostrophic flow $v_0=r^2$ for different Ekman numbers (as labelled). 
(a) We compare $\left(v_0\right)_\mathrm{vol.}$ (solid lines) and $\left(v_0\right)_\mathrm{Wang}$ (dotted lines) to the geostrophic flow $v_0=r^2$, plotted as a black line. 
(b) Difference in the bulk: $(\left(v_0\right)_\mathrm{vol.}-\left(v_0\right)_\mathrm{Wang})/Ek^{1/2}$ versus $r$.
(c) Difference in the boundary layer: $(\left(v_0\right)_\mathrm{vol.}-\left(v_0\right)_\mathrm{Wang})/\Ek^{1/2}$ versus $(r-1)/\Ek^{1/4}$.
}
\label{f.CompWang}
\end{figure}

For this comparison, we use a simple geostrophic flow $v_0(r) = r^2$. 
In figure~\ref{f.CompWang}\textit{a}, we plot both $\left(v_0\right)_\mathrm{vol.}$ and $\left(v_0\right)_\mathrm{Wang}$ for this geostrophic flow  for different values of the Ekman number. 
Although  some differences are visible when the Ekman number is moderately small ($\Ek \approx 10^{-4}$), these differences vanish when $\Ek$ is asymptotically small.
In figure~\ref{f.CompWang}\textit{b} and \textit{c}, we demonstrate that they are $O(\Ek^{1/2})$ in the bulk and $O(\Ek^{1/4})$ in the side wall boundary layer, respectively.   
We indeed see that  for small Ekman numbers 
$(\left(v_0\right)_\mathrm{vol.}-\left(v_0\right)_\mathrm{Wang})/\Ek^{1/2}$ becomes constant in the bulk, while $(\left(v_0\right)_\mathrm{vol.}-\left(v_0\right)_\mathrm{Wang})/\Ek^{1/4}$ converges   close to the wall to a function of the boundary layer variable $(r-1)/\Ek^{1/4}$ independent of $\Ek$.

\bibliographystyle{jfm}
\bibliography{geobib}
	
\end{document}